\documentclass{article}

\usepackage{arxiv}

\usepackage[utf8]{inputenc} 
\usepackage[T1]{fontenc}    
\usepackage{hyperref}       
\usepackage{url}            
\usepackage{booktabs}       
\usepackage{amsfonts}       
\usepackage{nicefrac}       
\usepackage{microtype}      
\usepackage{lipsum}		
\usepackage{graphicx}
\usepackage{subfig}

\title{New Method for Silicon Sensor Charge Calibration Using Compton Scattering}

\author{
  Patrick~McCormack\thanks{wpmccormack@lbl.gov}\\
  Physics Department\\
  UC Berkeley\\
   \And
 Maurice~Garcia-Sciveres \\
  Physics Division\\
  Lawrence Berkeley National Lab\\
   \And
 Timon~Heim \\
  Physics Division\\
  Lawrence Berkeley National Lab\\
   \And
 Benjamin~Nachman \\
  Physics Division\\
  Lawrence Berkeley National Lab\\
   \And
 Magne~Lauritzen \\
  Department of Physics and Technology\\
  University of Bergen\\
}

\begin{document}
\maketitle

\begin{abstract}
In order to cope with increasing lifetime radiation damage expected at collider experiments, silicon sensors are becoming increasingly thin.  To achieve adequate detection efficiency, the next generation of detectors may have to operate with thresholds below 1000 electron-hole pairs.  The readout chips attached to these sensors should be calibrated to some known external charge, but there is a lack of traditional sources in this charge regime.  We present a new method for absolute charge calibration based on Compton scattering.  In the past, this method has been used for calibration of scintillators, but to our knowledge never for silicon detectors.  Here it has been studied using a 150 micron thick planar silicon sensor on an RD53A readout integrated circuit.
\end{abstract}


\section{Introduction}
\subsection{Motivation and requirements for absolute charge calibration}
\label{need}
Silicon pixel detectors are an important technology for experiments at the Large Hadron Collider (LHC) at CERN due to their radiation hardness and readout rate~\cite{LHC, PERF-2007-01, CMS, ALICE}.  Typically, they comprise the innermost elements of the detectors and are used for tracking and vertexing.  Hybrid pixel and monolithic active pixel silicon (MAPS) detectors are currently prominent silicon technologies used by CMS, ATLAS, and ALICE~\cite{cmos1, cmos2, ALICEITS}.

The use of increasingly thin sensors in silicon trackers is fueled by multiple reasons: lower mass and higher radiation tolerance for hybrid detectors~\cite{radhard} and the increasing use of monolithic detectors~\cite{ALICEITS, pixelref}.  Furthermore, radiation damage decreases the charge collection for a given thickness, and the need to push silicon sensor radiation tolerance results in readout electronics sensitive to ever lower signals.  Currently, the thicknesses of sensors used in these experiments' detectors range from 200 $\mu$m to 285 $\mu$m, depending on the experiment and location in the detector~\cite{Aad:2008zz, ATLAS-TDR-19, CMStracker}.  Planned upgrades to detectors will use even thinner sensors.  For example, the ATLAS ITk hybrid pixel detector ~\cite{Collaboration:2285585, ATL-PHYS-PUB-2019-014} will include 100~$\mu$m thick planar sensors with an end-of-life average charge per minimum ionizing particle (MIP) of 6000 electron-hole (e-h) pairs or fewer at normal incidence.  The equivalent value for the ALICE ITS~\cite{ALICEITS} inner layer monolithic sensors, which will be 50~$\mu$m thick, is 3400 e-h pairs or fewer.  Throughout this paper, a conversion of 3.6 eV per e-h pair will be assumed~\cite{eh_energy}.

To achieve maximum detection efficiency for lower-energy hits, operating thresholds are set to be as low as possible while avoiding noise occupancy.  Throughout this paper, the ``threshold'' can be interpreted as the energy deposit which would be detected 50\% of the time.  A slightly smaller energy deposit would be detected less frequently and a slightly larger deposit more frequently.  Pixel detectors are typically operated with thresholds of order 10 ENC, where ENC is the Equivalent Noise Charge obtained by converting the amplifier noise to units of input charge.  Current pixel detectors in ATLAS and CMS typically operate with a hit threshold between 2000 e-h pairs and 5000 e-h pairs (7.2 - 18.0 keV), depending on the detector layer and year~\cite{atlaspixop, cmspixop, cmspixop2}.  For the next generation of pixel detectors, thresholds may be in the range of 300 e-h pairs to 1700 e-h pairs (1 - 6 keV).

Silicon tracking detectors typically have an internal charge injection circuit for in situ tuning and threshold verification.  However, this injection circuit must be validated with an external source to be considered accurate.  The external source makes an absolute calibration possible by depositing a known amount of charge in the sensor.  Throughout this paper, the units for the nominal charge injected by the internal circuit will be denoted as \~e in order to distinguish from true charge, which will be denoted as ``e-h pairs''.  By performing a calibration, a function is obtained that converts nominal injected charge to true charge: $\mathcal{F}:$~\~e~$\rightarrow$~e-h pairs.  Beyond absolute charge scale calibration, external sources may be necessary to check the uniformity of the injection circuit across a front-end.  Additionally, there can be uncertainties in the charge collection from the sensor after irradiation, and any method other than generation of a known signal in the sensor itself will be subject to those uncertainties.

Historically, such absolute calibrations have relied on techniques such as X-ray absorption from fluorescence~\cite{fluorcalib, fluorcalib2, fluorcalib3} or the absorption of radiation from radioisotope sources with known gamma- or X-ray peaks~\cite{sourcecalib}.  Studies are also done using the energy loss of minimum ionizing particles (MIPs) to provide qualitative comparisons and find outliers~\cite{mipcalib}.  For a variety of reasons, these techniques are not always adequate below 6 keV:

\begin{itemize}
\item Irradiating a sensor with X-rays between 1 and 6 keV is possible using the K$_{\alpha}$~and K$_{\beta}$~lines of the elements Na through Mn~\cite{xraydb}.  However, this requires the use of expensive equipment and the use of multiple elemental sources.
\item There is no typically used radioisotope source with an energy peak below that of $^{55}$Fe's 5.9 keV X-ray line.  Other typical radioisotopes would be $^{241}$Am (59.5 keV line) and $^{109}$Cd (22 and 25 keV lines)~\cite{radioisotope}.
\item A MIP signal is not monochromatic and does not lend itself to characterization of the detector threshold behavior or the response function.  MIP calibrations are also often not practical.
\end{itemize}

With these problems in mind, there are several practical goals and advantages of the calibration method explained in this paper, which is based on Compton scattering.  It should provide access to energies between 1 and 6 keV by using commonly available monochromatic photon sources.  The calibration signal itself will be monochromatic, but a single setup will provide access to a continuous spectrum of energies.  Additionally, its setup should be lower cost than that of an X-ray fluorescence setup.

\subsection{Compton scattering for calibration}
\label{compton}
\begin{figure}[!htb]
  \centering
  \subfloat[]{\includegraphics[width=0.45\linewidth]{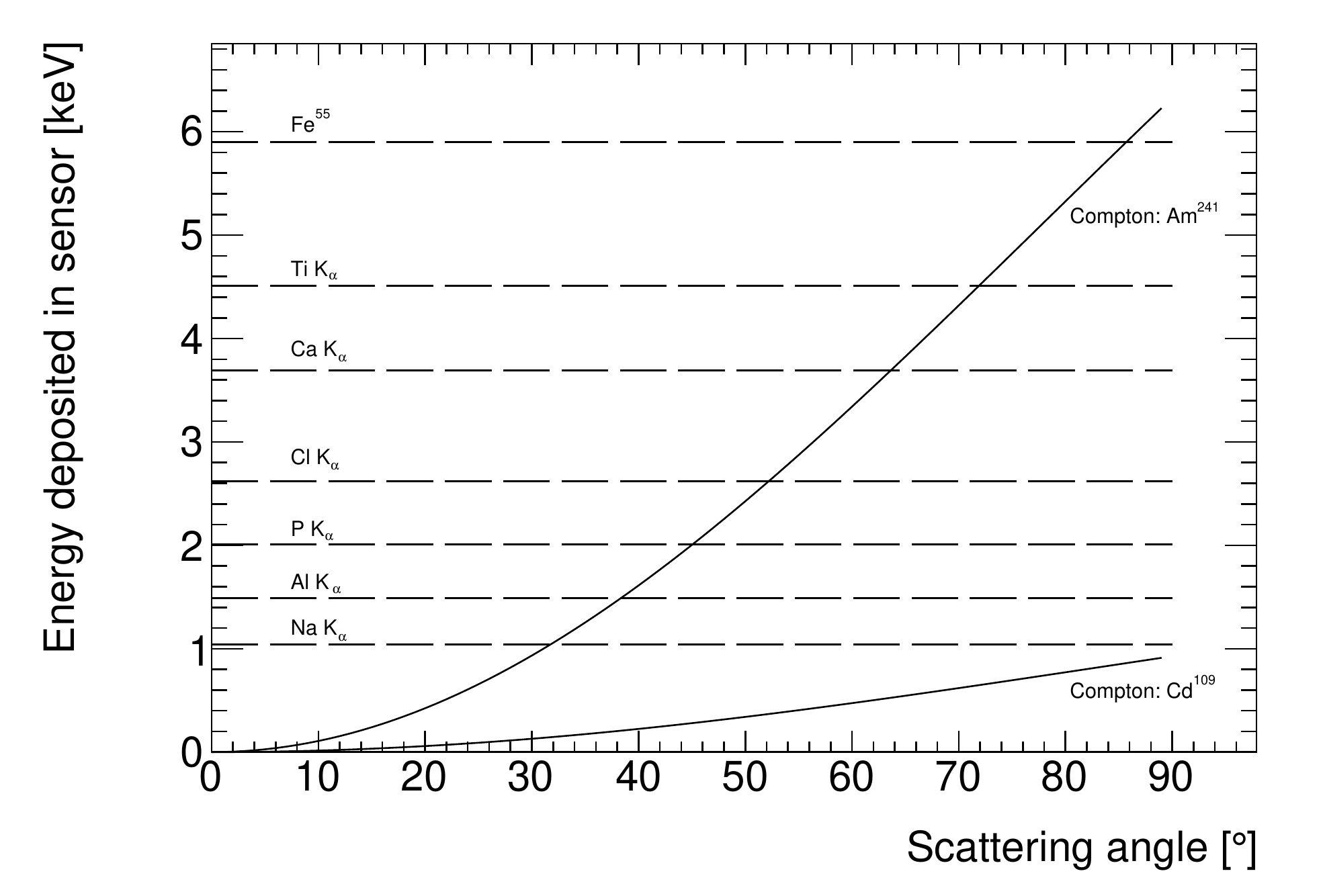}}
  \qquad
  \subfloat[]{\includegraphics[width=0.45\linewidth]{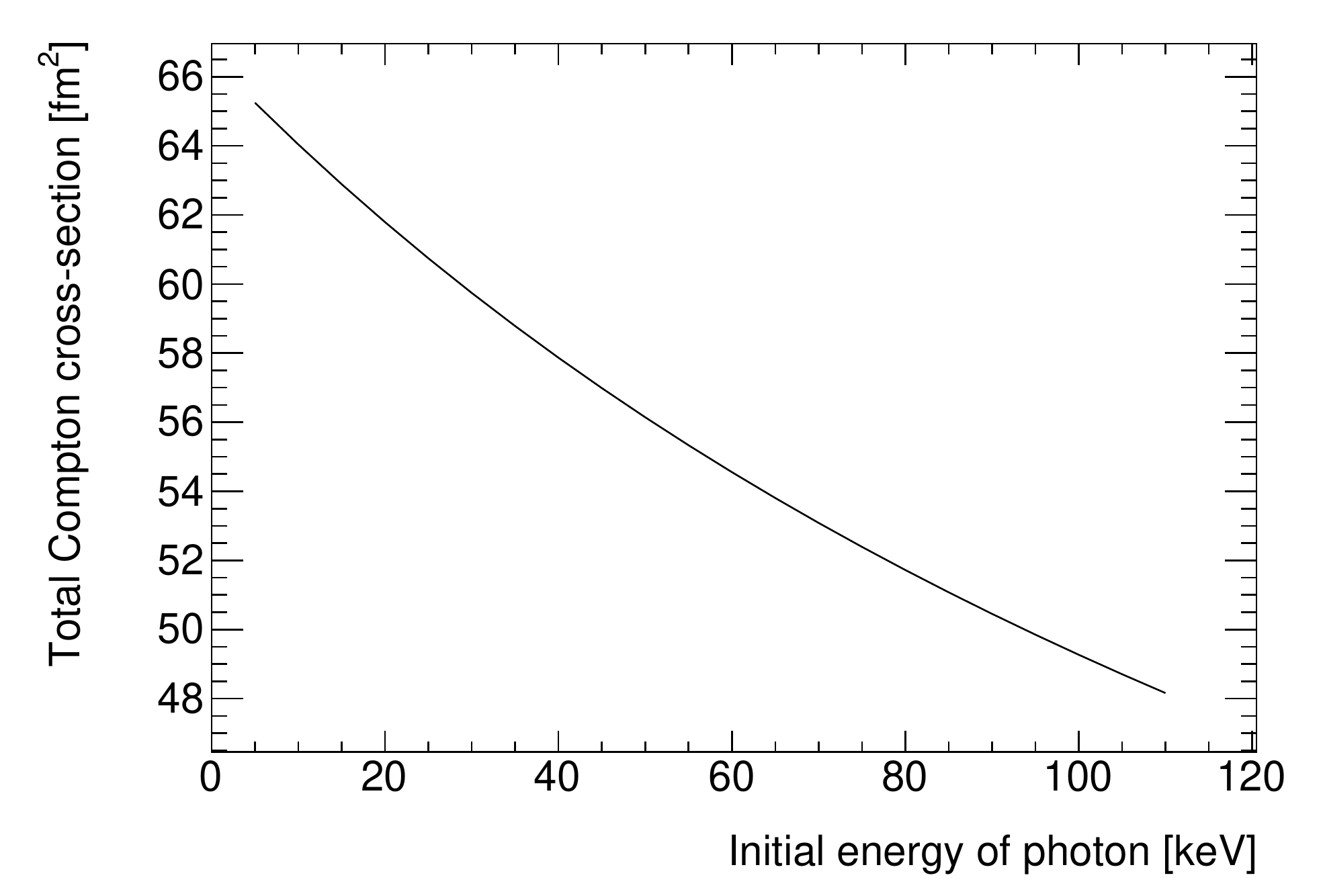}}
  \caption{(a) Solid lines: energy imparted to a silicon sensor as a function of the Compton-scattering angle.  This is just $E_{\gamma} - E_{\gamma '}$, with $E_{\gamma '}$ from Equation~\ref{comp}.  Lines shown for an $^{241}$Am~source (59.5 keV) and a $^{109}$Cd source (22 keV).  Dashed lines: select radioisotope and X-ray fluorescence lines.  These are just for comparison and are not scattering-angle dependent.  (b) Total cross-section for Compton scattering as a function of the incoming photon energy.}
  \label{energyplot}
\end{figure}

The Compton scattering of photons off the electrons within a silicon sensor can be used to deposit 1 - 6 keV of energy into the sensor.  The energy of a photon after Compton scattering off an electron is~\cite{compton}:
\begin{equation}
\label{comp}
E_{\gamma '} = \frac{E_{\gamma}}{1 + (E_{\gamma}/m_e c^2)(1 - \cos \theta)},
\end{equation}
where $E_{\gamma '}$ is the scattered photon's energy, $E_{\gamma}$ is the initial energy, and $\theta$ is the scattering angle of the photon.  The scattered electron acquires the energy $E_{\gamma} - E_{\gamma '}$, and this energy will be promptly reabsorbed as a localized energy deposit within the sensor.  For example, if photons from an $^{241}$Am source (peak at 59.5 keV) are scattered at angles in the range $5^{\circ}$ to $90^{\circ}$, then between 0.03 and 6.2 keV will be deposited.  A plot of the energy deposited in the sensor as a function of the scattering angle for photons from an $^{241}$Am source is shown as the solid line in Figure~\ref{energyplot}~(a).  The energies of the radioisotope $^{55}$Fe and of select X-ray fluorescence lines are also shown for comparison, as are the energies of Compton-scattered $^{109}$Cd photons.  Depending on the physical setup used, larger scattering angles could be used to access higher energies.  Alternatively, a higher energy radioisotope could be used to produce the initial photons.

The differential cross-section for Compton scattering is given by the Klein-Nishina formula~\cite{KNform}.  The total cross-section for the Compton scattering of a 59.5 keV photon is $5.47 \times 10^{-25}$~cm$^{2}$.  This varies depending on the energy as shown in Figure~\ref{energyplot}~(b).

To know the deposited energy, an X-ray detector can be placed to detect photons scattering into a small angular region.  Any photons detected by the X-ray detector at a given scattering angle will have deposited approximately the same energy in the sensor.  The total charge of each deposit associated with a particular scattering angle will not be known exactly due to the Doppler broadening of the Compton peak~\cite{doppler1, doppler2, doppler3}.  This broadening of the scattered energy peak is due to the fact that the electrons within the sensor are not at rest.  For incident photons with energy of 59.5 keV, the FWHM of the peak of photons scattered at $90^{\circ}$ is expected to increase by 0.6 keV~\cite{res1, res2}.  Because of this effect, the Compton scattering method is best suited for an absolute calibration aggregated over many pixels.

\begin{figure}[!htb]
  \centering
  \includegraphics[width=0.7\linewidth]{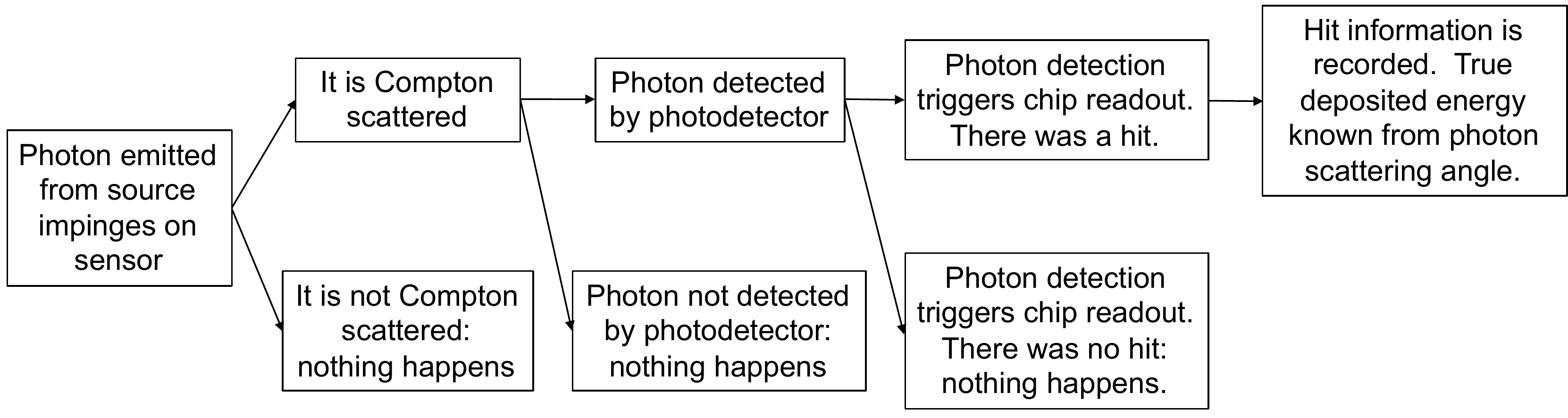}
  \caption{Flowchart of the Compton-scattering method, from the emission of a photon to data-collection.}
  \label{dataflow}
\end{figure}

The X-ray detector can be used to trigger data readout from a chip attached to the silicon sensor.  In this way, a photon detected by the X-ray detector and an energy deposit in the silicon can be correlated.  The position of the X-ray detector can be adjusted so that the calibration can be performed over a continuous range of energies.  A flowchart of the processes behind Compton scattering method is presented in Figure~\ref{dataflow}.  One can trace the steps that lead to data collection once the initial photon has been emitted.

\section{Instruments and setup}
\subsection{Device under study}
\label{device}
\begin{figure}[!htb]
  \centering
  \includegraphics[width=0.3\linewidth]{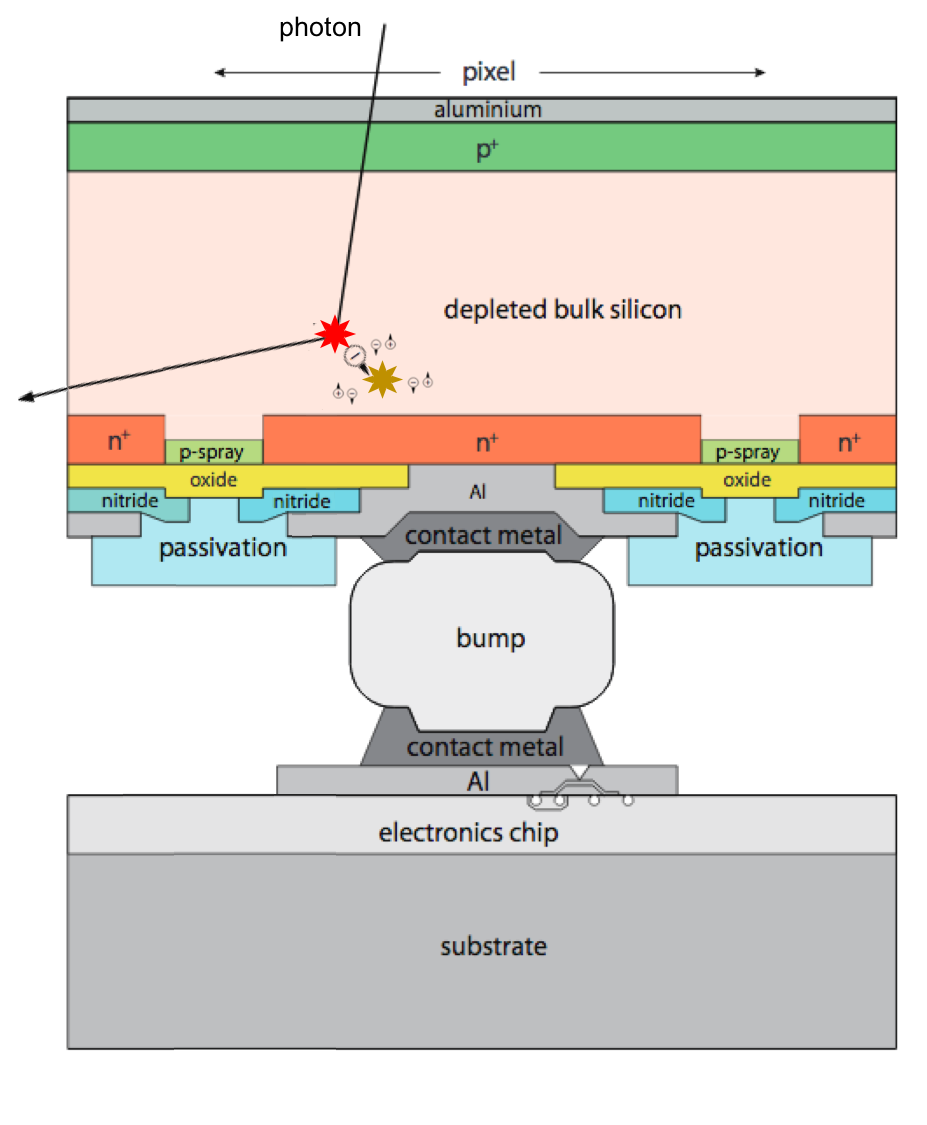}
  \caption{Example geometry of a hybrid pixel detector.  A passive silicon sensor is bump-bonded to a readout chip, which serves as the active electronics~\cite{pixelref}.  An incident photon is shown Compton scattering off of an electron at point indicated by the red star.  The electron deposits its energy within the sensor at the point indicated by the gold star, creating electron-hole pairs that drift within the depleted bulk.}
  \label{hybrid}
\end{figure}

A hybrid pixel detector was used in the development of this calibration technique, and the geometry of such a detector is shown in Fig~\ref{hybrid}~\cite{pixelref}.  An example of a Compton scattering leading to a charge deposit is also illustrated in the figure.

The sensor used to study this calibration method was a 150~$\mu$m silicon sensor manufactured by MPG-HLL~\cite{HLL}, which had 100x25~$\mu$m pixels.  The sensor was bump-bonded onto an RD53A readout integrated circuit (RD53A)~\cite{rd53a}, which served as a readout chip.  The total size of the chip is 20.0 mm by 11.6 mm, but only the differential front-end of the RD53A used, which leads to an active area of 6.8mm x 9.2mm.  Readout and tuning were performed by using the YARR software framework~\cite{yarr}.  The RD53A and its sensor are mounted on a custom-designed single chip Printed Circuit Board (PCB).  The sensor, chip, and PCB unit will be referred to as the ``silicon module assembly''.

The front end of the RD53A is a charge sensitive amplifier (CSA) with constant reset~\cite{rd53a}.  The height of the analogue pulse the readout chip receives is proportional to the total charge deposited in the sensor.  There is a threshold on readout, such that if the pulse does not rise above the threshold, then there is no readout.  This is the same threshold introduced in Section~\ref{need}.  The pulse is digitized as a time over threshold (ToT) reading, which is the integer number of clock cycles that the pulse stays above threshold.  The RD53A has a 40 MHz chip clock.  If the pulse were to barely exceed the threshold and dissipate before the next clock cycle, the ToT would be 1.  This counts as detection.  The ToT response for a given charge is manipulated by adjusting both the threshold and the signal fall time (or return to baseline).  The settings for these parameters are based on input from the internal charge injection circuit.  For the RD53A, ToT is stored as a 4-bit output in the datastream.  Examples of the ToT response to injected charge for a particular tuning will be shown in Section~\ref{test1}.

\begin{figure}[!htb]
  \centering
  \includegraphics[width=0.55\linewidth]{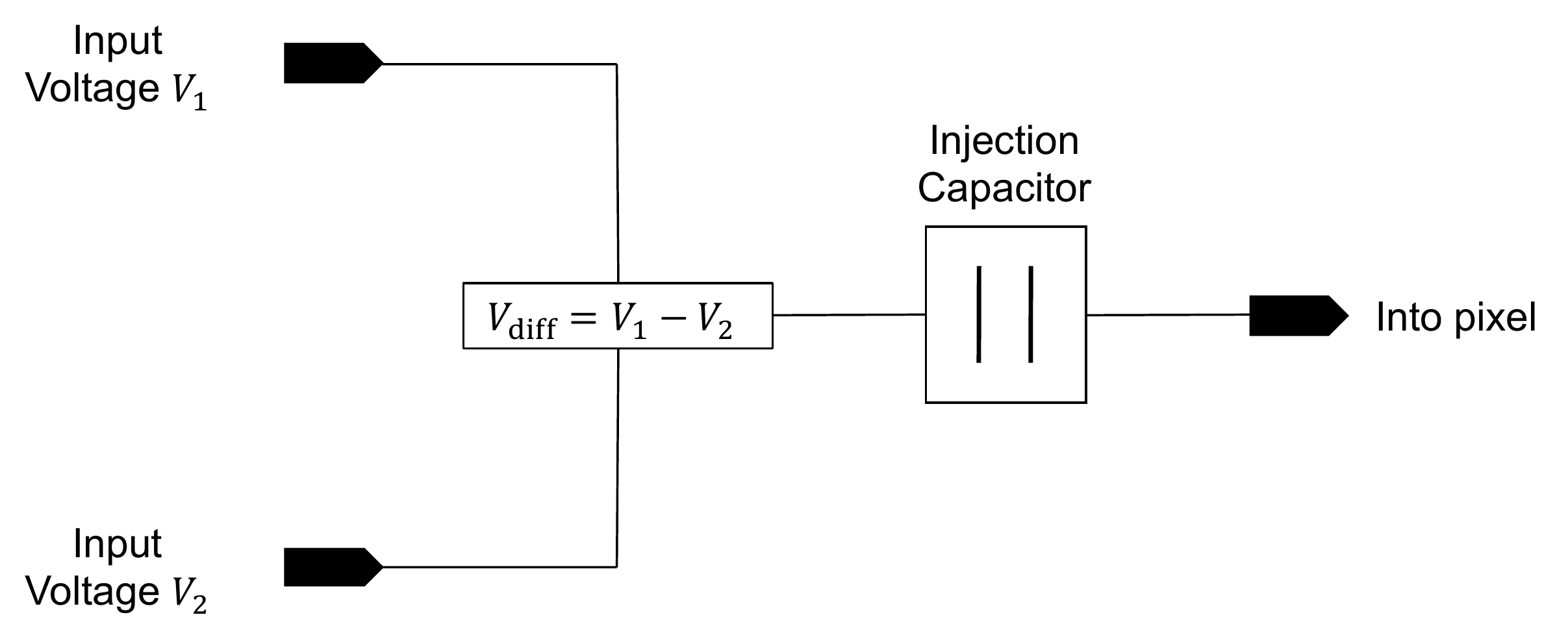}
  \caption{Simple schematic of the calibration injection circuit in each pixel of the RD53A readout integrated circuit.  Two input voltages are used to charge a capacitor, which then releases its charge into the pixel.  This circuit is in parallel to the actual sensor.}
  \label{inj_circuit}
\end{figure}

Every pixel in the RD53A has a charge-injection circuit, which is used for a pixel-by-pixel tuning.  A simplified schematic of this circuit is shown in Figure~\ref{inj_circuit}.  Two input DC voltages can be selected for one terminal of a capacitor.  Switching between these two voltages results in a charge proportional to the voltage difference being injected into the CSA.  Most silicon pixel detectors will similarly have an internal injection circuit, and Compton scattering should provide the data needed calibrate this circuit to an absolute charge scale.

By adjusting both the signal threshold and the signal fall rate, the 4-bit ToT output can be adjusted to cover a variety of energy ranges.  The RD53A is capable of tuning to thresholds below 500 \~e.  For the differential front-end, the functional relationship between input charge and ToT is non-linear and pixel dependent.

In typical running, RD53A can self-trigger readout based on sensor hits.  However, it can also be externally triggered.  The data acquisition system (YARR) includes settings and scans based on such external triggering.

\subsection{In-lab setup}
\label{instruments}

To detect scattered photons, an Amptek X-123 x-ray spectrometer with a 5mm diameter CdTe detector element was used~\cite{x123}.  The spectrometer has auxiliary ports that allow for the output of a simple logic pulse whenever a photon is detected.  This was the external trigger for the RD53A.  For reference, a flowchart of the steps leading to data readout was presented in Figure~\ref{dataflow}.

\begin{figure}[!htb]
  \centering
  \subfloat[]{\includegraphics[trim=50 0 200 50, clip, width=0.45\linewidth]{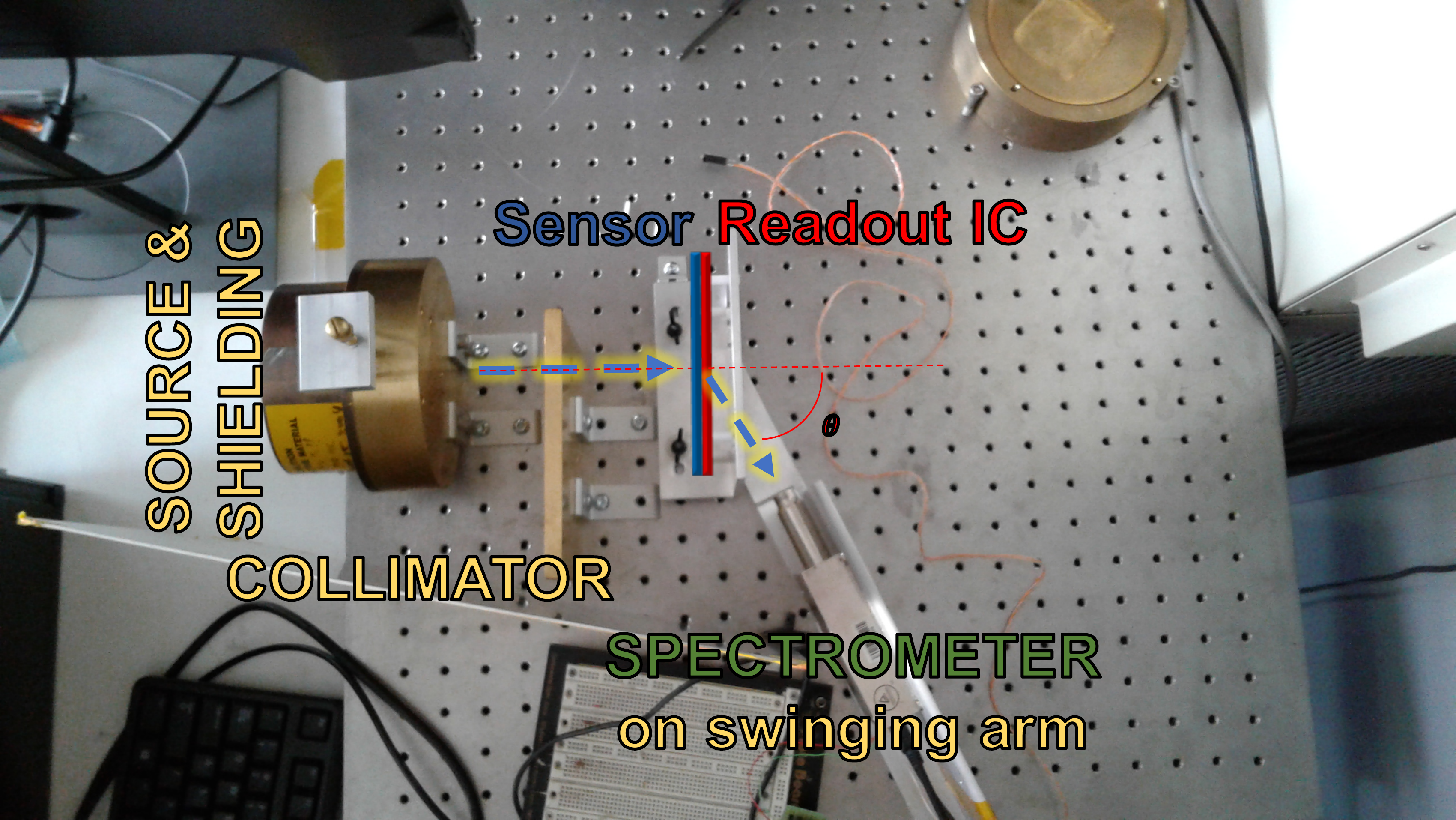}}
  \qquad
  \subfloat[]{\includegraphics[width=0.45\linewidth]{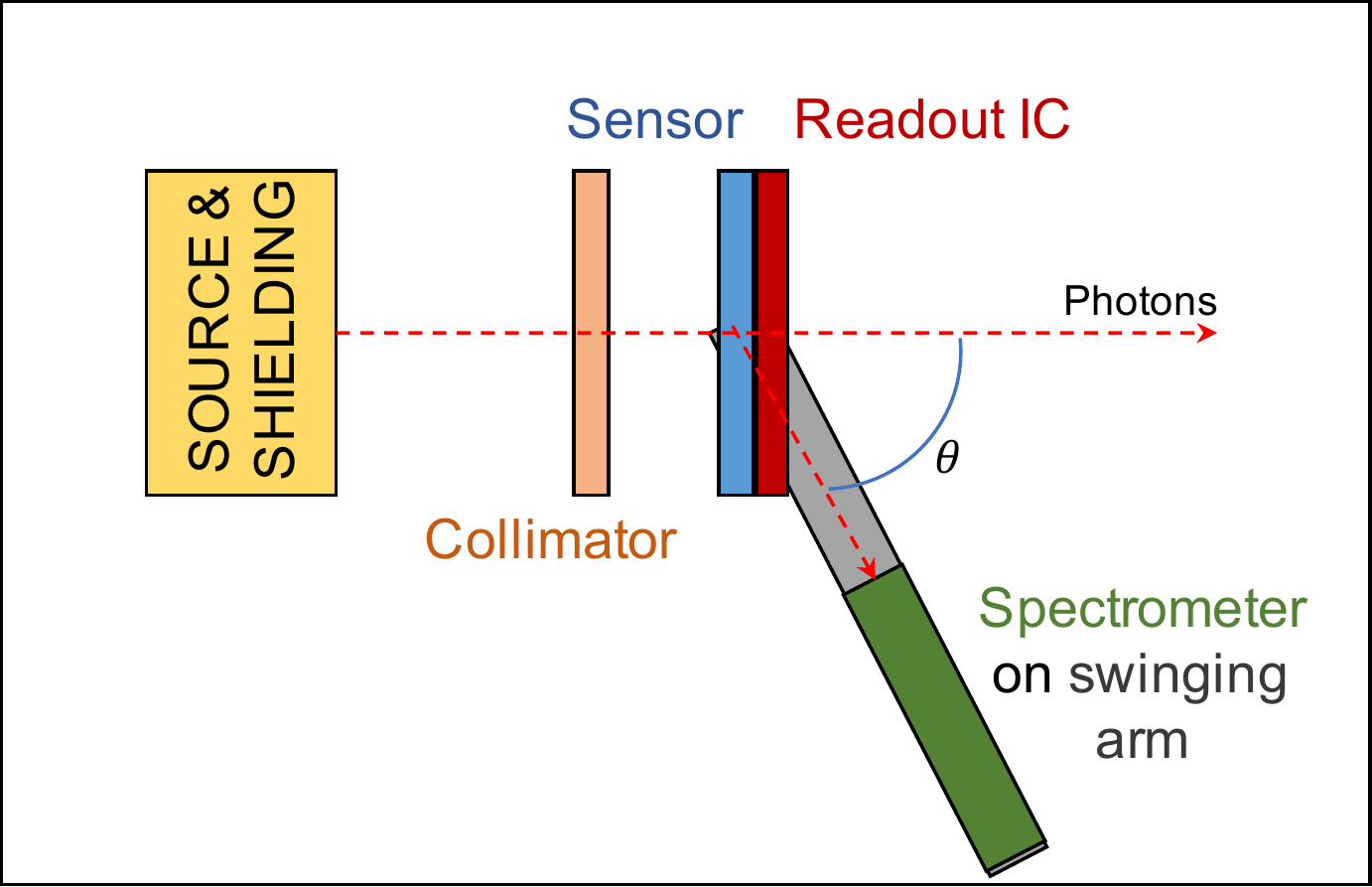}}
  \caption{(a) Picture of the experimental setup.  59.5 keV photons are emitted from an Americium source through a hole in surrounding brass shielding, pass through a brass collimator, scatter off of the silicon sensor bump bonded to the readout integrated circuit, and are detected by the spectrometer. (b) Diagram of setup.}
  \label{setuppic}
\end{figure}

A picture and diagram of the experimental setup from above are presented in Figure~\ref{setuppic}.  The equipment is laid out on an optical bench with a 1'' x 1'' grid of holes.  An $^{241}$Am source with activity of 100 mCi is placed on one end of the grid; there is a 3mm aperture in a piece of brass shielding in front of the source that creates a beam of photons.  The $^{241}$Am spectrum has a 59.5 keV peak, with a measured FWHM of 0.55 keV.  The photons pass through an additional brass collimator with a diameter of 3mm before falling on the silicon module assembly.  The spectrometer is mounted on a single arm pivot with rotation point below the sensor, which allows the angle of the scattered-photon acceptance window to be easily adjusted.  During data taking, the setup is covered by a box to block out ambient light which could be a background to the silicon sensor and spectrometer.  The scattering of photons from the collimator, non-sensor parts of the module assemby, and support structures are a background for the spectrometer.


\subsection{Spectrometer readout}
\label{trigger}
\begin{figure}[!htb]
  \centering
  \includegraphics[width=0.55\linewidth]{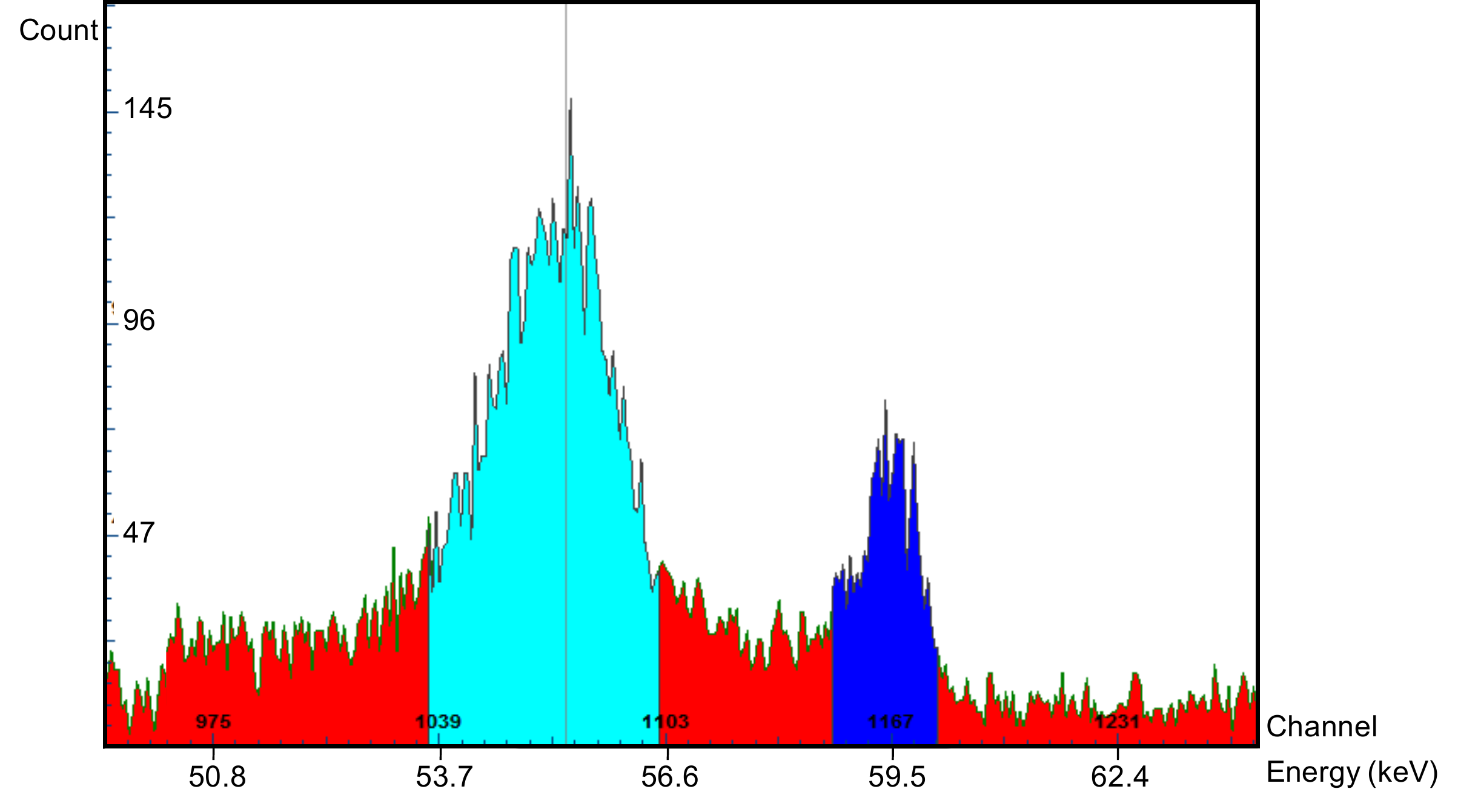}
  \caption{Example photon spectrum for Compton scattering of photons at an angle of 71.6$^\circ$.  The peak associated with Compton scattered photons is in light blue; the peak associated with Thompson scattered photons is in dark blue.}
  \label{spectrom_spectrum}
\end{figure}

\begin{figure}[!htb]
  \centering
  \includegraphics[width=0.55\linewidth]{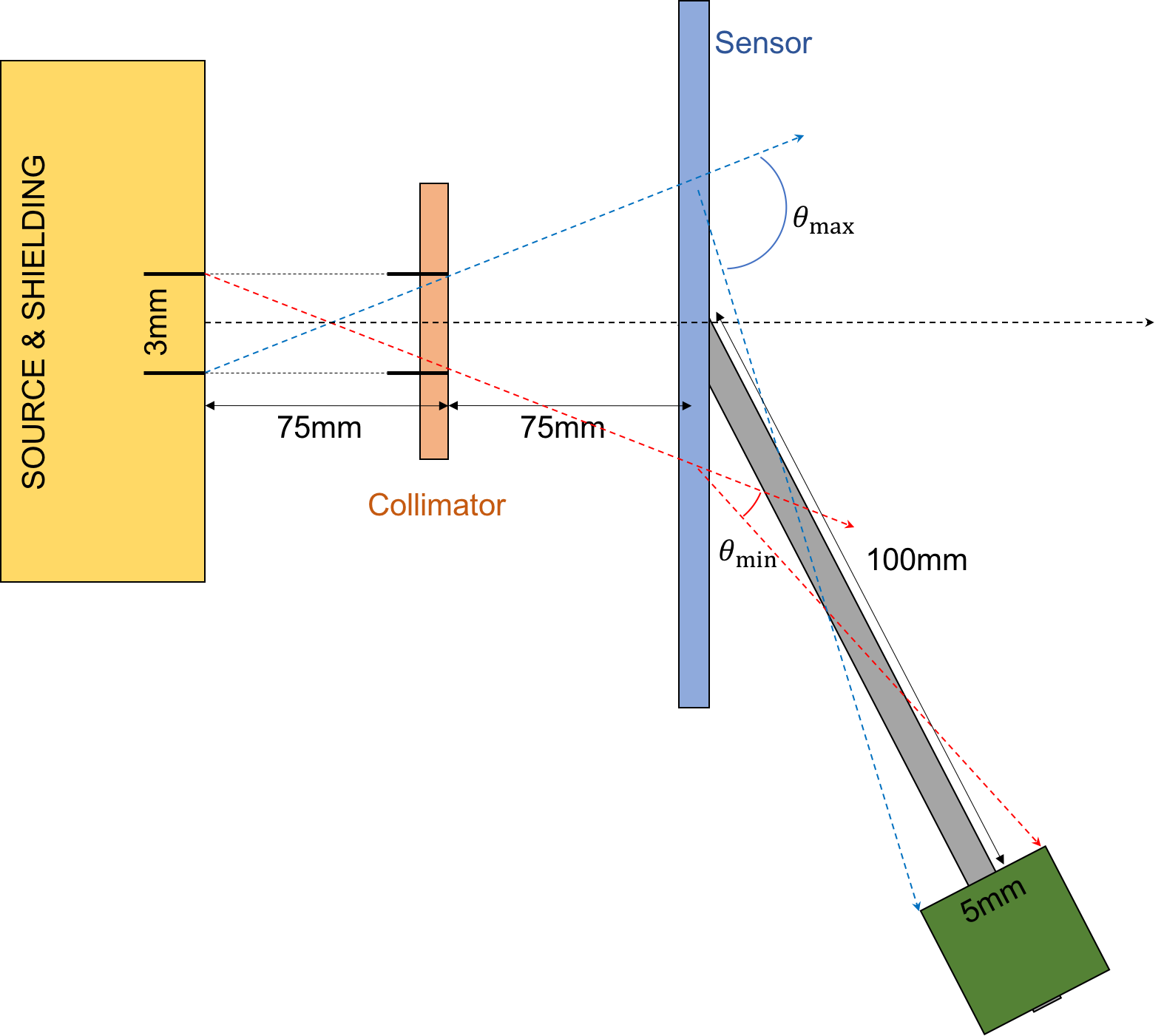}
  \caption{Diagram of the scattering scenarios that lead to the geometric broadening of the Compton peak.  Distances are not shown to scale here in order to make small distances and angles visible.  The minimum scattering angle that the spectrometer will see, $\theta_{\mathrm{min}}$, is shown by the red dashed line, and the maximum angle, $\theta_{\mathrm{max}}$, is shown by the blue dashed line.}
  \label{angles}
\end{figure}

An example of the spectrometer's readout is shown in Figure~\ref{spectrom_spectrum}.  The spectrum was acquired at an angle of 71.6$^\circ$~relative to the beamline.  Here, two peaks can be seen, the larger one in light blue is the peak associated with Compton scattering.  Because of the scattering angle, this peak is centered on an energy value of 55.1 keV.  The smaller peak on the right is caused by the Thompson scattering of photons off of nuclei.  These photons lose little energy, and as such, the peak is at 59.5 keV.  The FWHM of the Compton peak, at 1.4 keV, is 2.5 times larger than that of the Thompson peak.  The broadening of the Compton scattering peak is caused by primarily by two factors.  One is the Doppler broadening, which was discussed in Section~\ref{compton}.  Second, slightly different angles of scattering are permitted due to the aperture sizes of the shielding, collimator, and spectrometer.  The geometries that lead to maximal and minimal scattering angles is portrayed in Figure~\ref{angles}.  Based around the center-line of the photon beam, the photons can be emitted within a distance of $\pm 1.5$~mm and they can be scattered within a distance of $\pm 4$~mm (this is somewhat restrained by the width of the front-end of the RD53A).  Over the 150 mm distance between the shielding aperture and the sensor, this represents an angular range of $\pm 2.1^\circ$ for the photons incident on the sensor.  Because the photon's incidence position can vary by $\pm 4$~mm with respect to the center-line, and the spectrometer aperture is $\pm 2.5$~mm, an additional angular smearing of $\pm 2.4^\circ$ is expected when the spectometer is positioned at a $60^\circ$~scattering angle.  Therefore, $\pm 4.5^{\circ}$ of different scattering angles are allowed at the spectrometer positions to be explored in the following calibration.  This effect alone would be expected to increase the FWHM by 0.8 keV.

\subsection{Resolution}
\label{practical}

\begin{figure}
  \centering
    \includegraphics[width=0.24\linewidth]{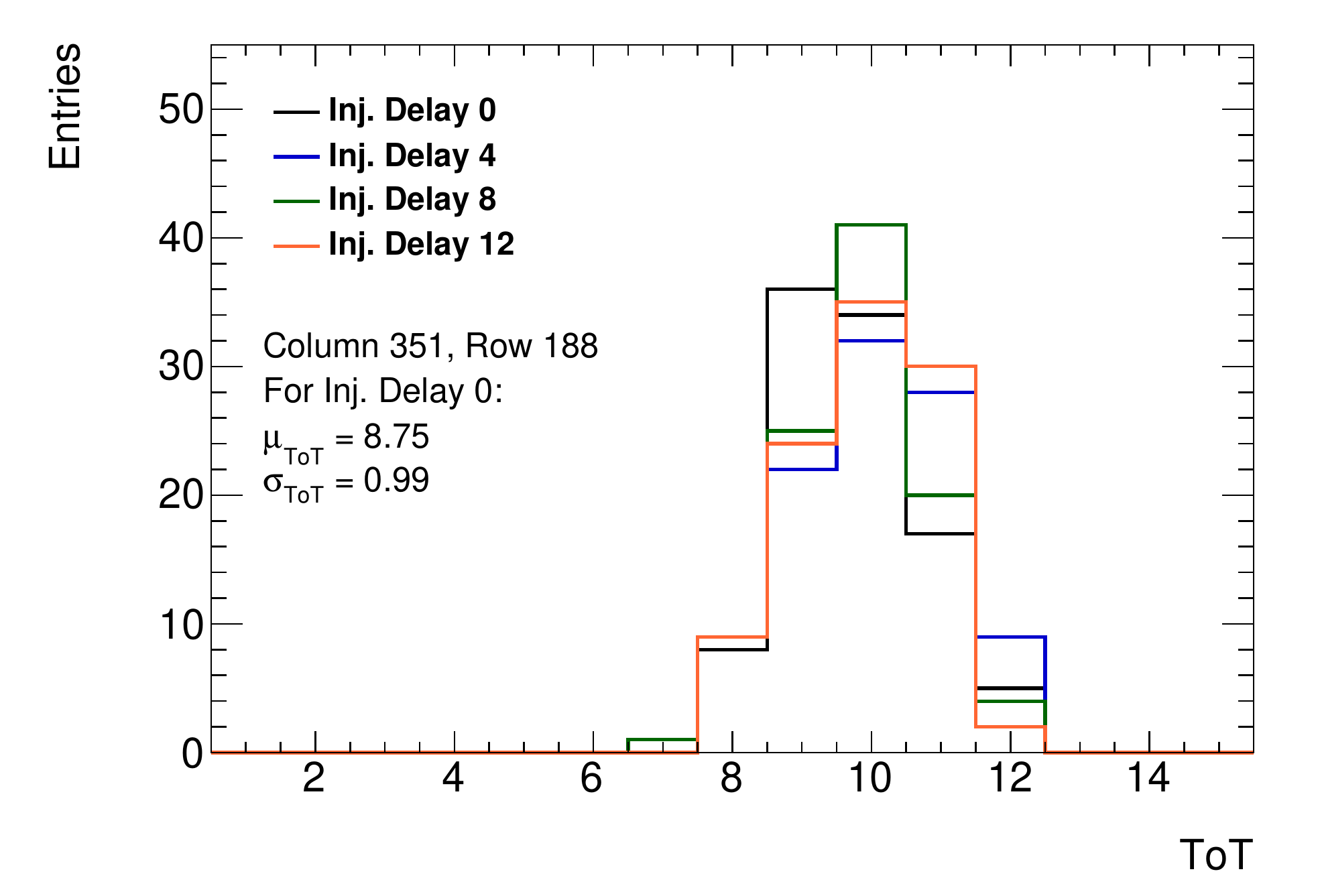}
    \includegraphics[width=0.24\linewidth]{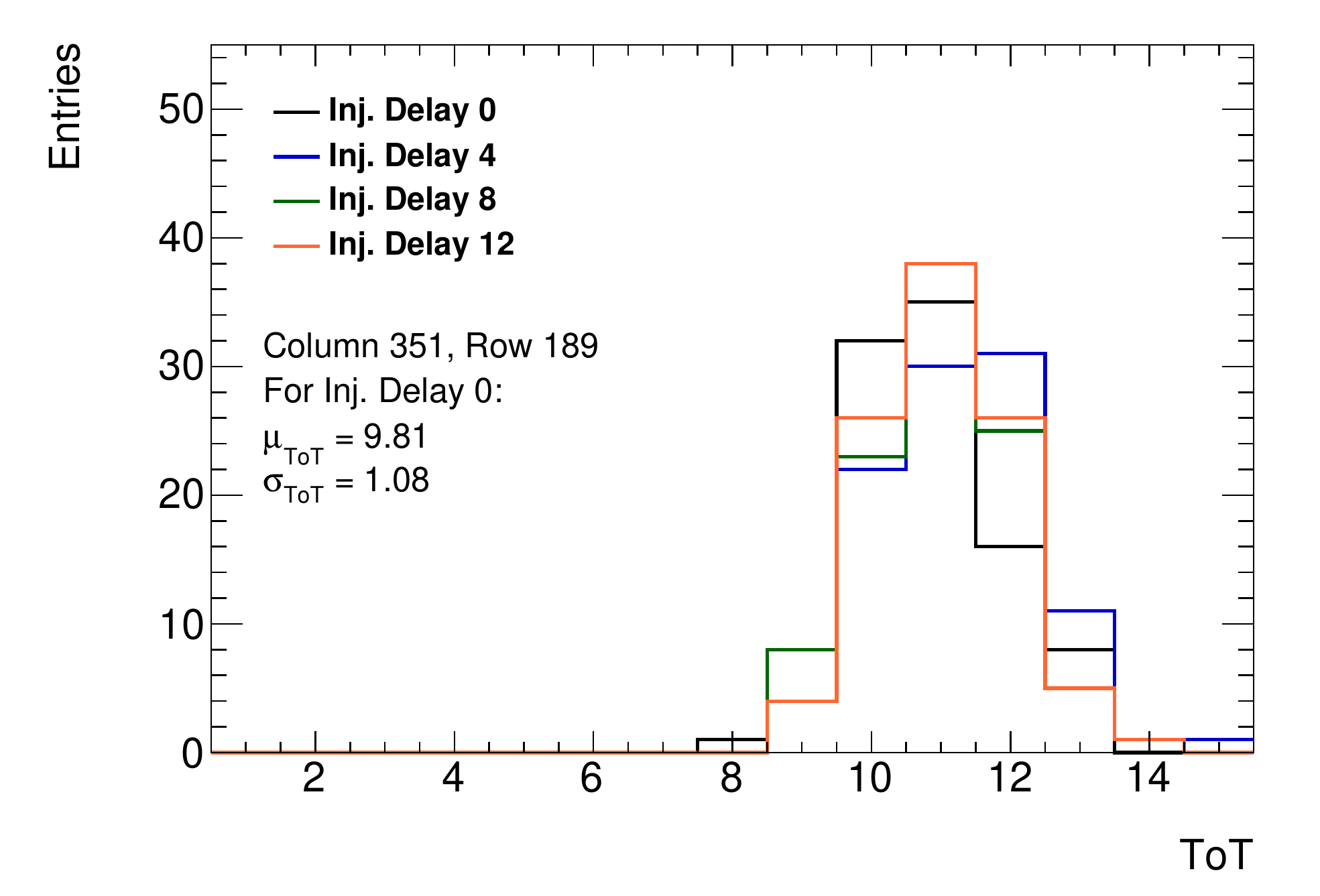}
   \includegraphics[width=0.24\linewidth]{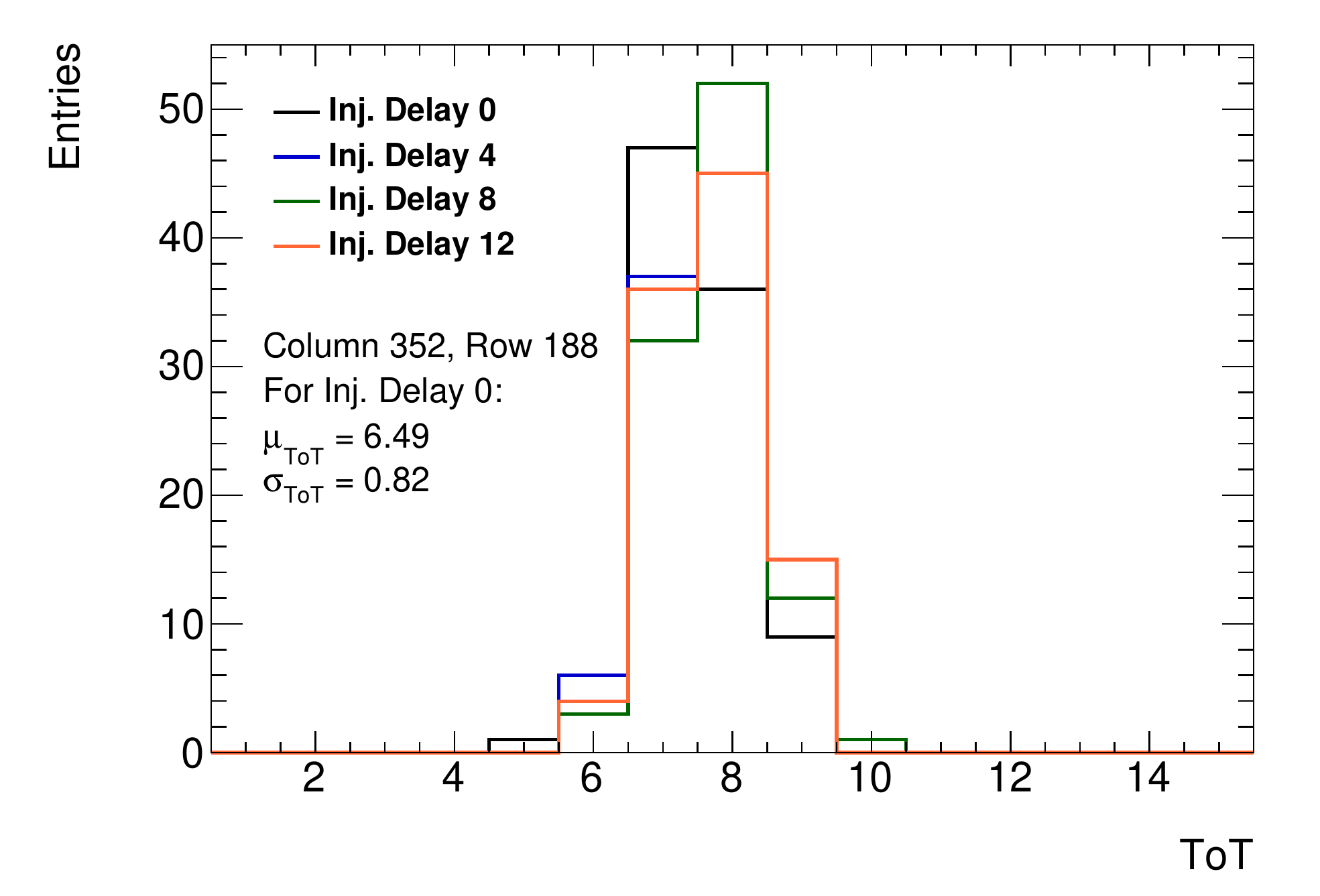}
   \includegraphics[width=0.24\linewidth]{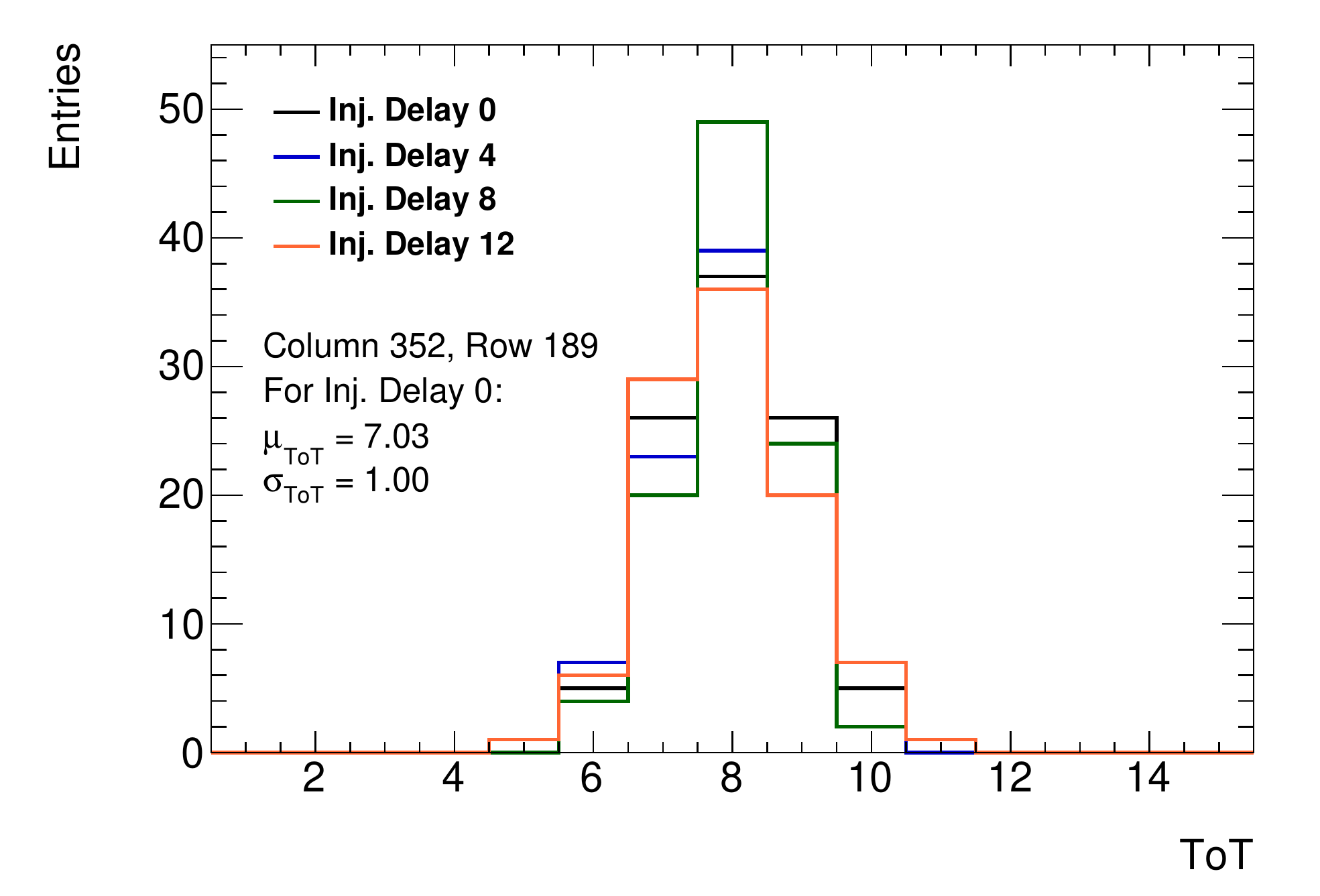}
  \caption{ToT distributions for 100 internal injections of 1000 \~e each into four different pixels.  The injections are additionally shown for four different injection fine delay settings for each pixel.  The fine delay affects when the internal injection occurs relative to the start of a clock cycle and is in units of 1/16 of a clock cycle.  Here it is used demonstrate the effect of a Compton scattering happening at different times relative to the start of a clock cycle.  The four pixels presented, in (column, row) format, are (351, 188), (351, 189), (352, 188), and (352, 189) from left to right.}
  \label{single_inject}
\end{figure}

There is an intrinsic dispersion in the ToT response of each pixel to a given charge, both one injected internally and one deposited externally.  In addition to manufacturing process variation making each pixel's transistors slightly different~\cite{manufacture}, charge deposits can happen at different times relative to the start of a clock cycle, influencing the number of cycles that will be counted, and there are non-uniformities in analog signal fall time.  If the same charge was injected into a single pixel multiple times, a distribution of ToT values with non-zero width would be expected.  This is demonstrated in Figure~\ref{single_inject}.  In this figure, four different pixels received 100 injections of 1000 \~e from the internal circuit, and each pixel has a different distribution of ToT values.  The figure also shows the distributions that occur for different injection fine delay settings.  The fine delay changes when the internal injection occurs relative to the start of a clock cycle, and here demonstrates the impact on resolution of charge deposit timing.

Due to the intrinsic variation mentioned above, each pixel will have a unique charge $\rightarrow$ ToT response function.  The knowledge of each pixel's function will be crucial for calibration.

\subsection{Tuning and observations}
\label{test1}
Tuning a chip is the process by which each pixel's threshold and charge to ToT response function is set.  Many internal injections are performed at the desired threshold value, and potential pixel settings are scanned, allowing for the selection of the correct settings.  Each pixel has 5 bits of threshold fine adjustment designed to allow the user to equalize the threshold of all pixels to the same value.

\begin{figure}[!htb]
  \centering
  \subfloat[]{\includegraphics[width=0.45\linewidth]{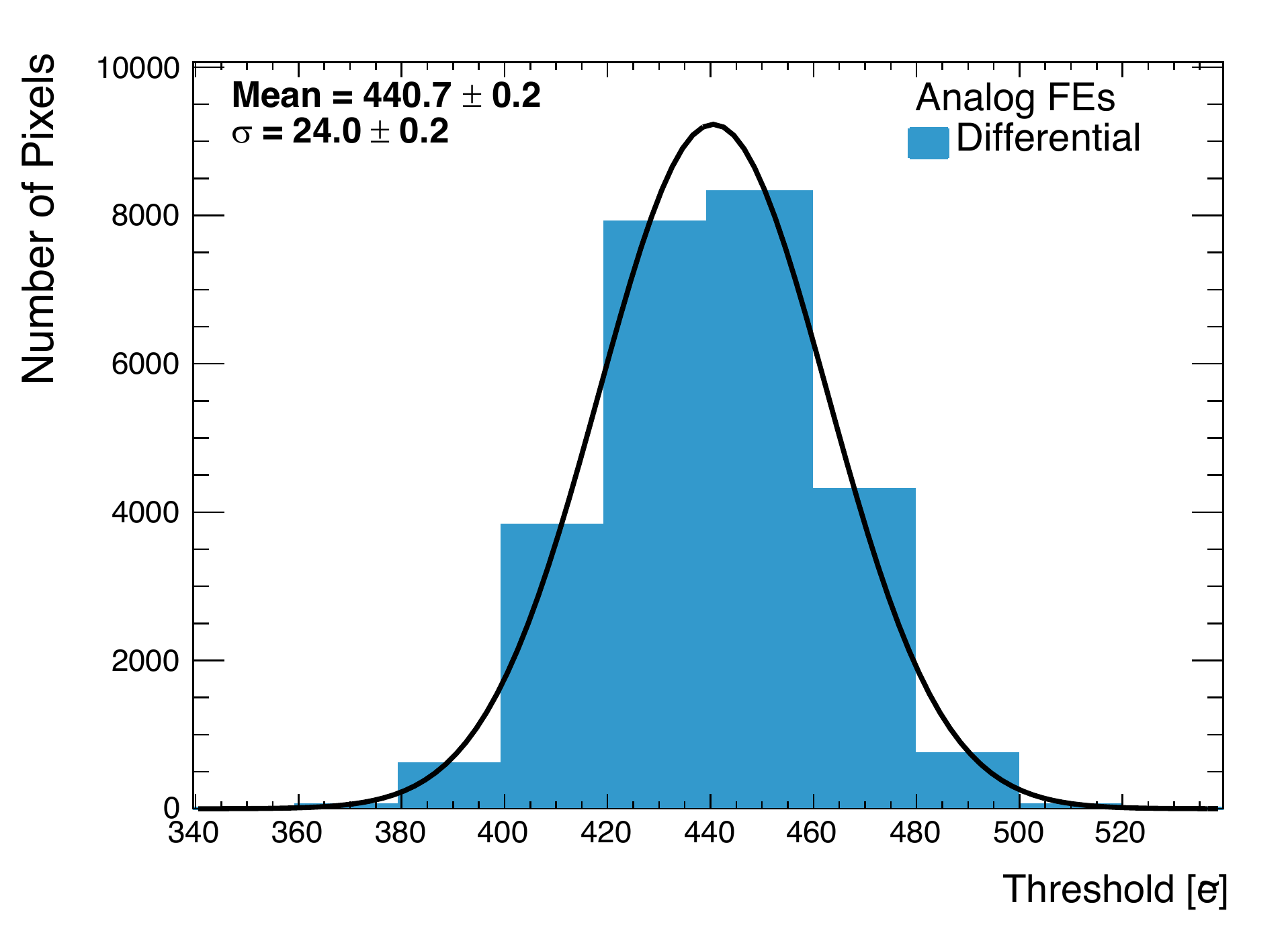}}
  \qquad
  \subfloat[]{\includegraphics[width=0.48\linewidth]{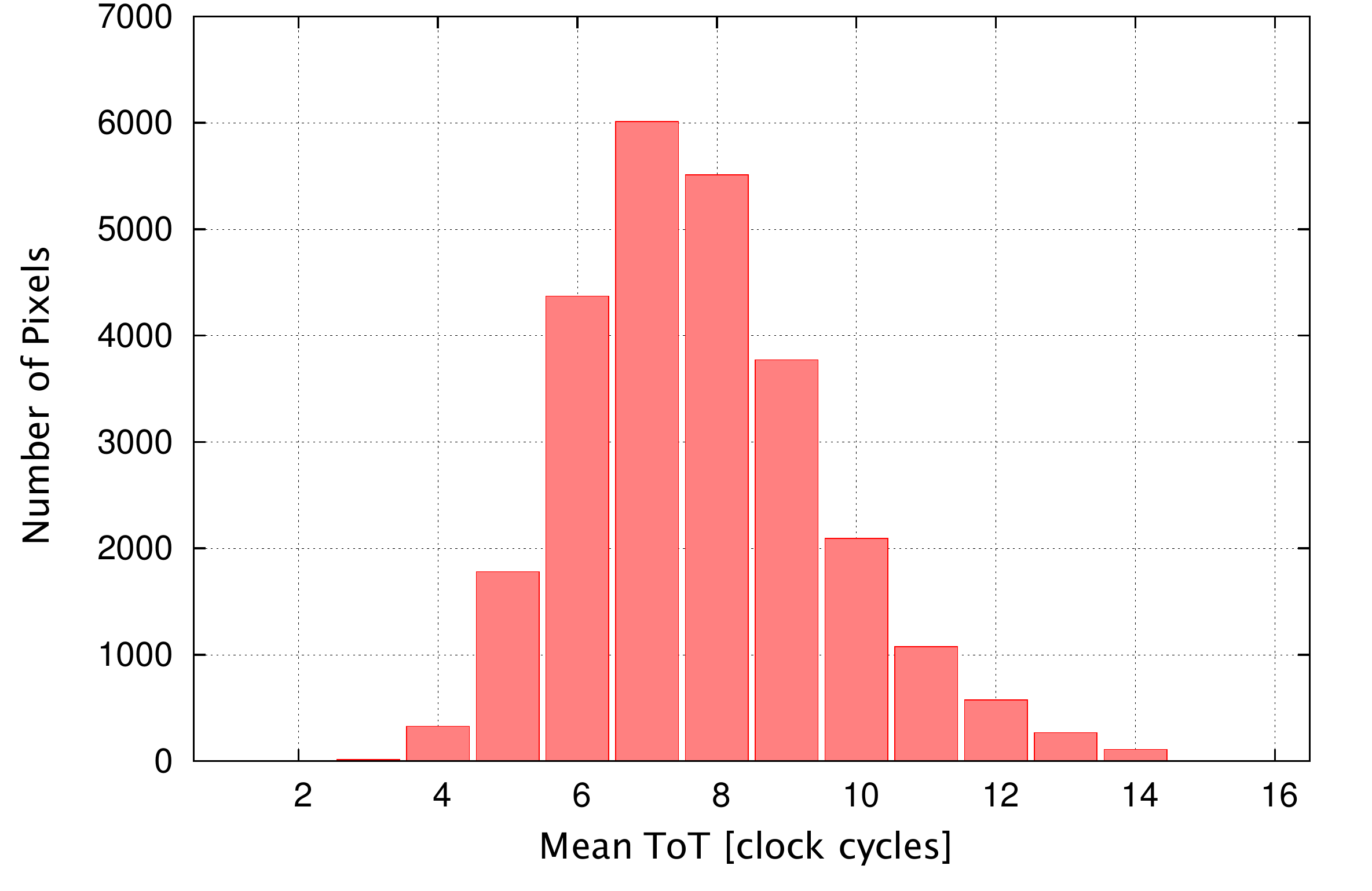}}
  \caption{(a) Histogram of the per-pixel threshold achieved in a chip tuning.  The average threshold achieved is 441 \~e.  (b) Histogram of all pixels' average ToT for repeated injections of 1000 \~e.  The average ToT across all pixels for such an injection is 8.76 clock cycles.}
  \label{tuning}
\end{figure}

For the purposes of a calibration using Compton scattering, the RD53A was tuned to have a threshold of 450 \~e and a slow signal fall rate.  A histogram of each pixel?s threshold after tuning is shown in Figure~\ref{tuning}~(a).  As discussed in Section~\ref{need}, this would mean that we would expect a hit to be recorded 50\% of the time if 450 \~e were to be injected into a random pixel.  The average pixel threshold achieved was 441 \~e, with a pixel-to-pixel standard deviation of 24 \~e.  Similarly, a histogram of all pixels' average ToT output due to a series of injections of 1000 \~e is shown in Figure~\ref{tuning}~(b).  Because of the slow return to baseline, the mean ToT is 8.8 clock cycles.  The pixel-to-pixel dispersion manifests here as a standard deviation of 1.8 clock cycles.  This can be compared to the single-pixel dispersion presented in Figure~\ref{single_inject}.

\begin{figure}[!htb]
  \centering
  \includegraphics[width=0.45\linewidth]{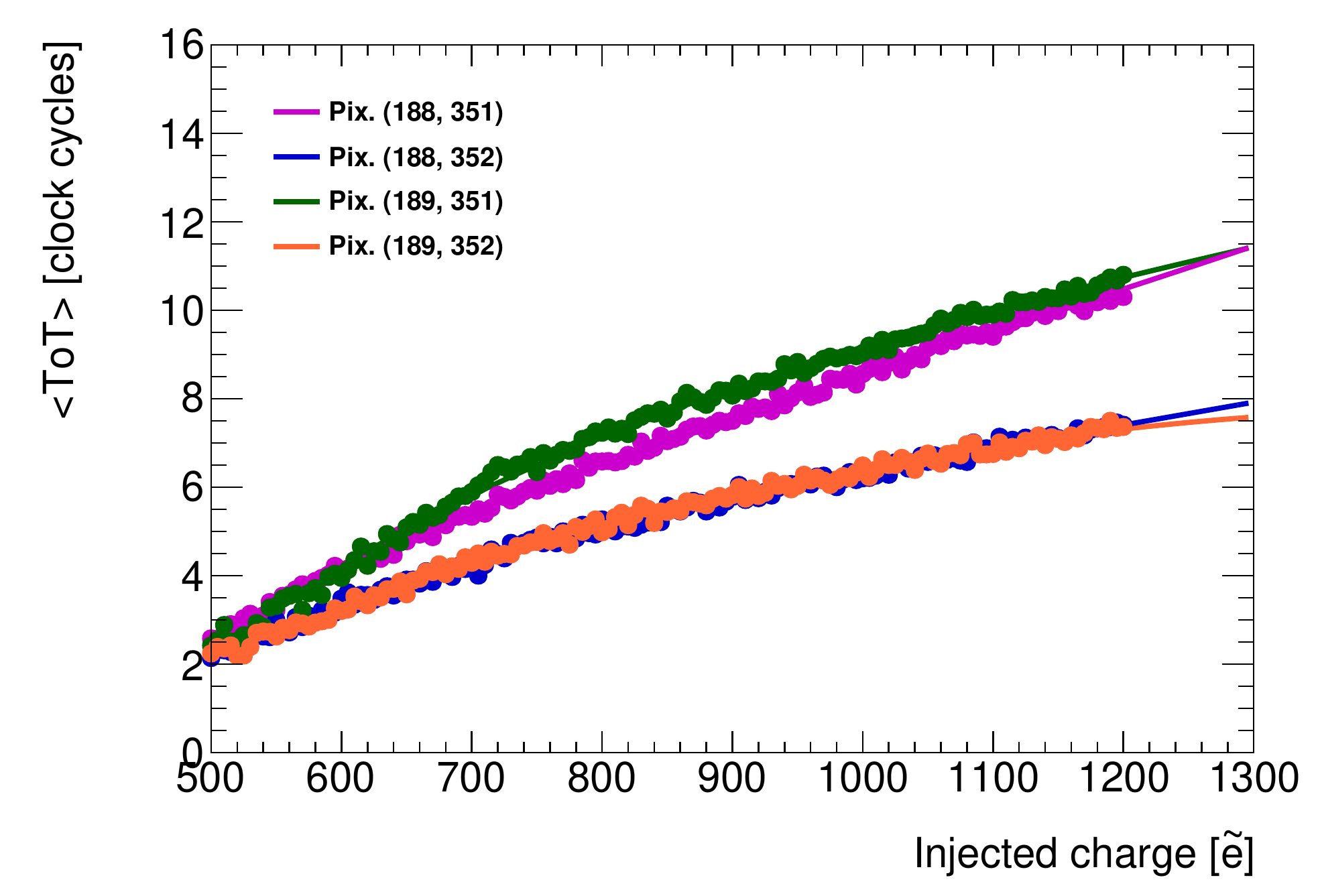}
  \caption{Four adjacent pixels' ToT response functions.  Points show the measured average ToT in the pixel for 100 charge injections each ranging from 500 \~e to 1245 \~e in steps of 5 \~e.  The solid lines show fits to a function of the form $a*\sqrt{x} + b + c*x + d*x^2$.}
  \label{totfunc}
\end{figure}

A unique ToT vs. injected charge response function exists for each pixel.  To approximate these functions, a series of charge injections are performed from 500 \~e to 1245 \~e in steps of 5 \~e.  Four adjacent pixels' ToT response functions are shown in Figure~\ref{totfunc}, illustrating the diversity of such functions found in the front-end.  Each pixel's plot is fit to a function of the form $a*\sqrt{x} + b + c*x + d*x^2$, such that each pixel has a unique set of $a,~b,~c,~d$.  In the Compton scattering runs, each hit is associated with an individual pixel.  Evaluating the above formula at the mean charge deposit associated with the photon's scattering angle yields the ``expected ToT'' of each hit.  This expected ToT will be compared to the actually observed ToT.

\subsection{Noise and backgrounds}
\label{noise}
\begin{figure}[!htb]
  \centering
  \subfloat[]{\includegraphics[width=0.45\linewidth]{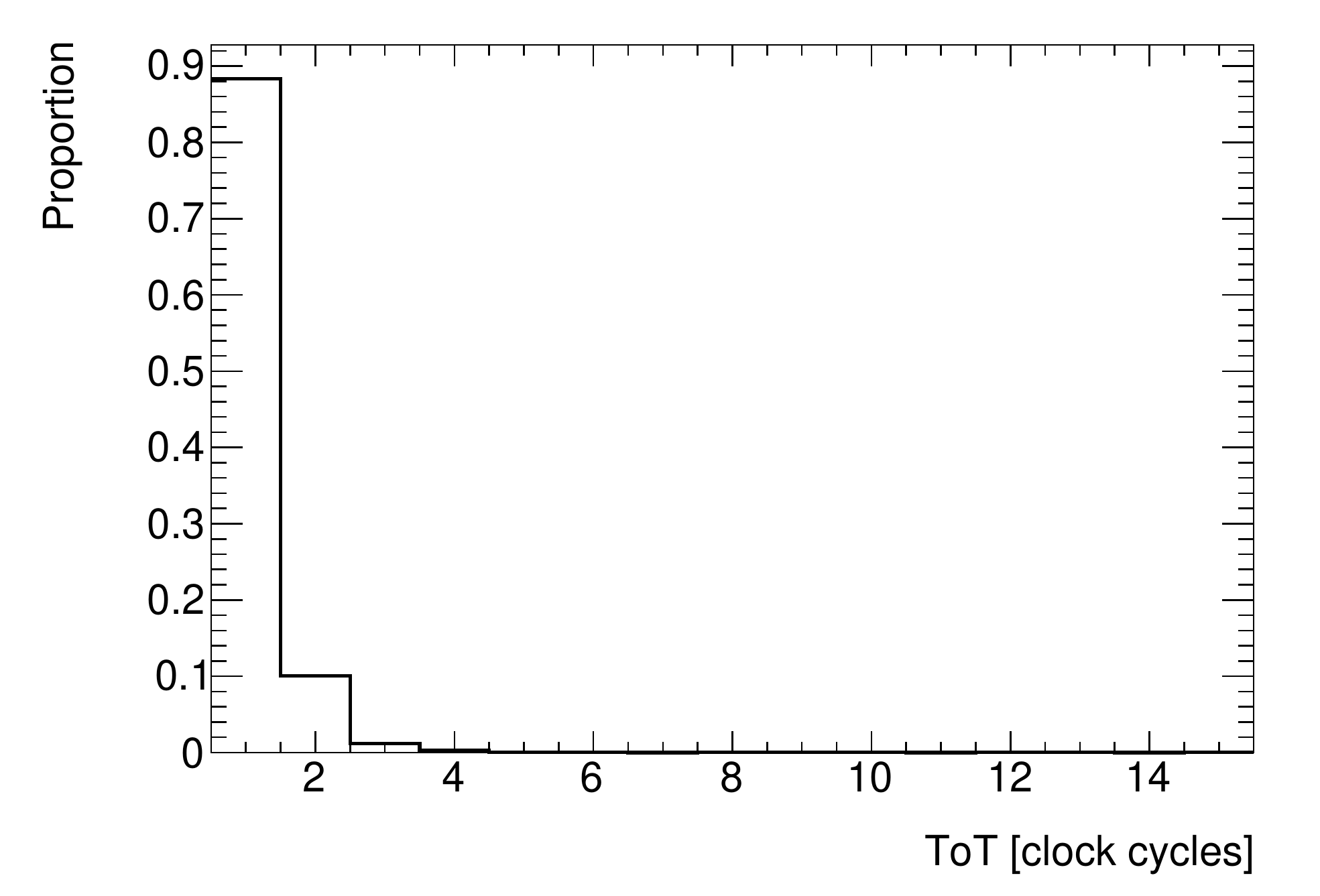}}
  \qquad
  \subfloat[]{\includegraphics[width=0.45\linewidth]{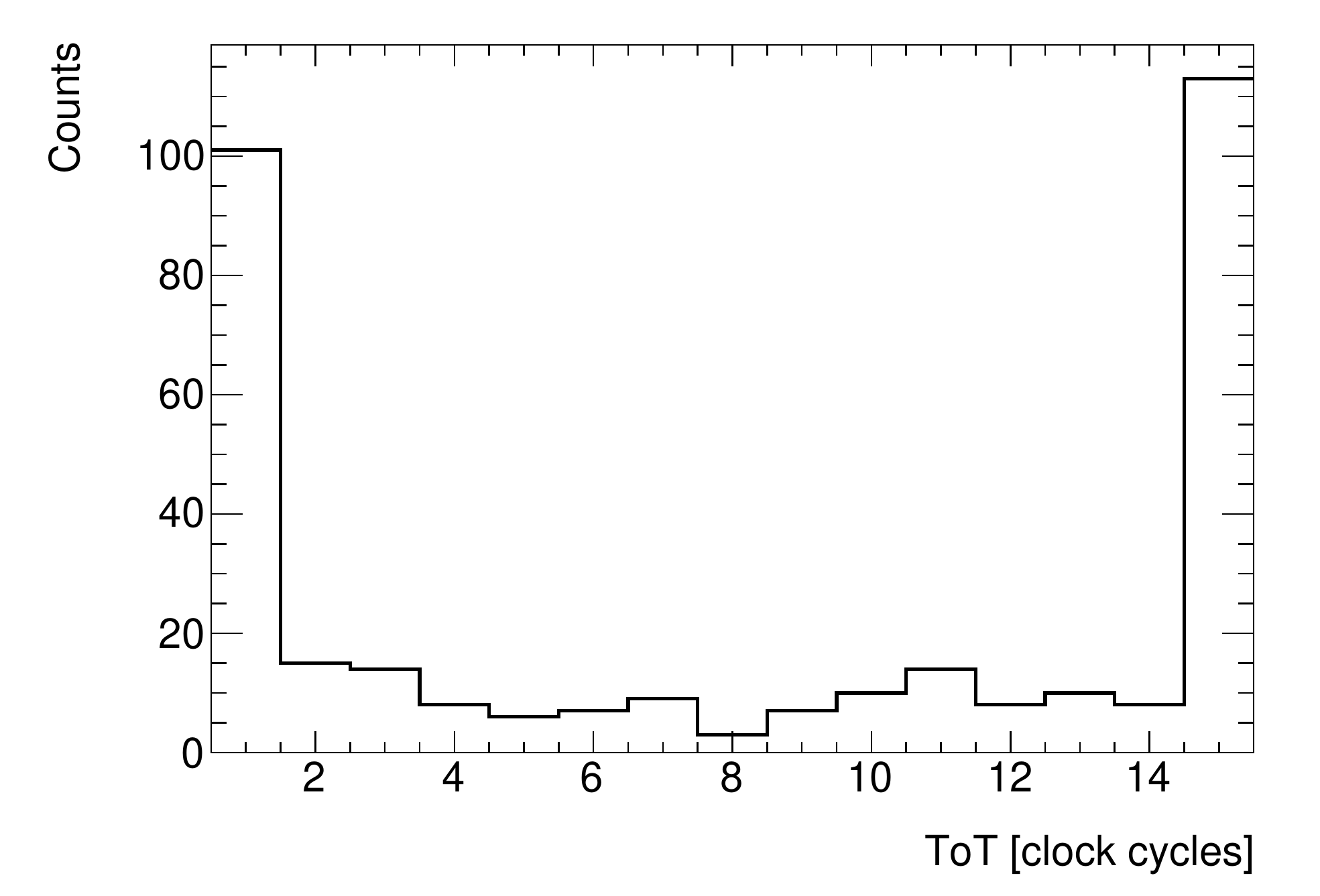}}
  \caption{(a) The hit distribution that results from recording data at random intervals.  There was no source of photons during this run, so these hits are considered to be pure noise.  The majority of noise hits have ToT = 1, with about 12\% having ToT = 2, and 1\% having a higher ToT.  (b) The hit distribution for random data taking with an $^{241}$Am source.  This distribution is a combination of noise and background.  Most hits with ToT = 1 are noise hits.  Hits will higher ToT tend to be background, which will be mostly Compton scatters into random directions here.}
  \label{n_and_b}
\end{figure}

As pixel threshold is lowered, noise fluctuations lead to recorded hits when no signal was present.  Most commonly, this will result in a hit with ToT of 1, with an approximately exponentially falling distribution.  The distribution of noise hits can be understood by recording data at random times with no photon source.  Such a distribution is shown in Figure~\ref{n_and_b}~(a).  It can be seen that the majority of noise hits have ToT = 1, with about 11\% having ToT = 2, and 1\% having a higher ToT.

A significant fraction of the photons detected by the spectrometer are not Compton scattered by an electron in the sensor, and so do not deposit any charge in the sensor.  Most of these photons have been scattered off of some other nearby object.  Most of the time this occurs, there will be no coincident hit in the RD53A, meaning that the trigger returns an empty event, as illustrated in Figure~\ref{dataflow}.  However, some fraction of the time there will be a coincident hit, which may be noise or a hit associated with a Compton scatter at a random angle.  To find the ToT distribution associated with this noise and background convolution, data is taken at random times with a photon source shining on the sensor.  The resulting distribution is shown in Figure~\ref{n_and_b}~(b).  Most of the hits with ToT = 1 here are noise, but hits with higher ToT tend to be from coincident X-ray hits.  About half of the background hits have ToT = 15, which corresponds to overflow.  These are hits where the pulse stays higher than threshold for a larger number of clock-cycles that can be encapsulated by the 4-bit ToT.  Background hits comprise about 70\% of this distribution, with noise hits as the other 30\%.

In the data-taking run described in Section~\ref{observe}, about 1/3 of hits that occurred were associated with hits with ToT 1 or 15.  Based on Figure~\ref{n_and_b}~(b), about 1/5 of hits with ToT between 2 and 14 are expected to be background.  In order to reduce background and noise contamination, the calibration will be performed using hits with ToT in the range [4,13].

\subsection{Observations}
\label{observe}
\begin{figure}[!htb]
  \centering
  \includegraphics[width=0.45\linewidth]{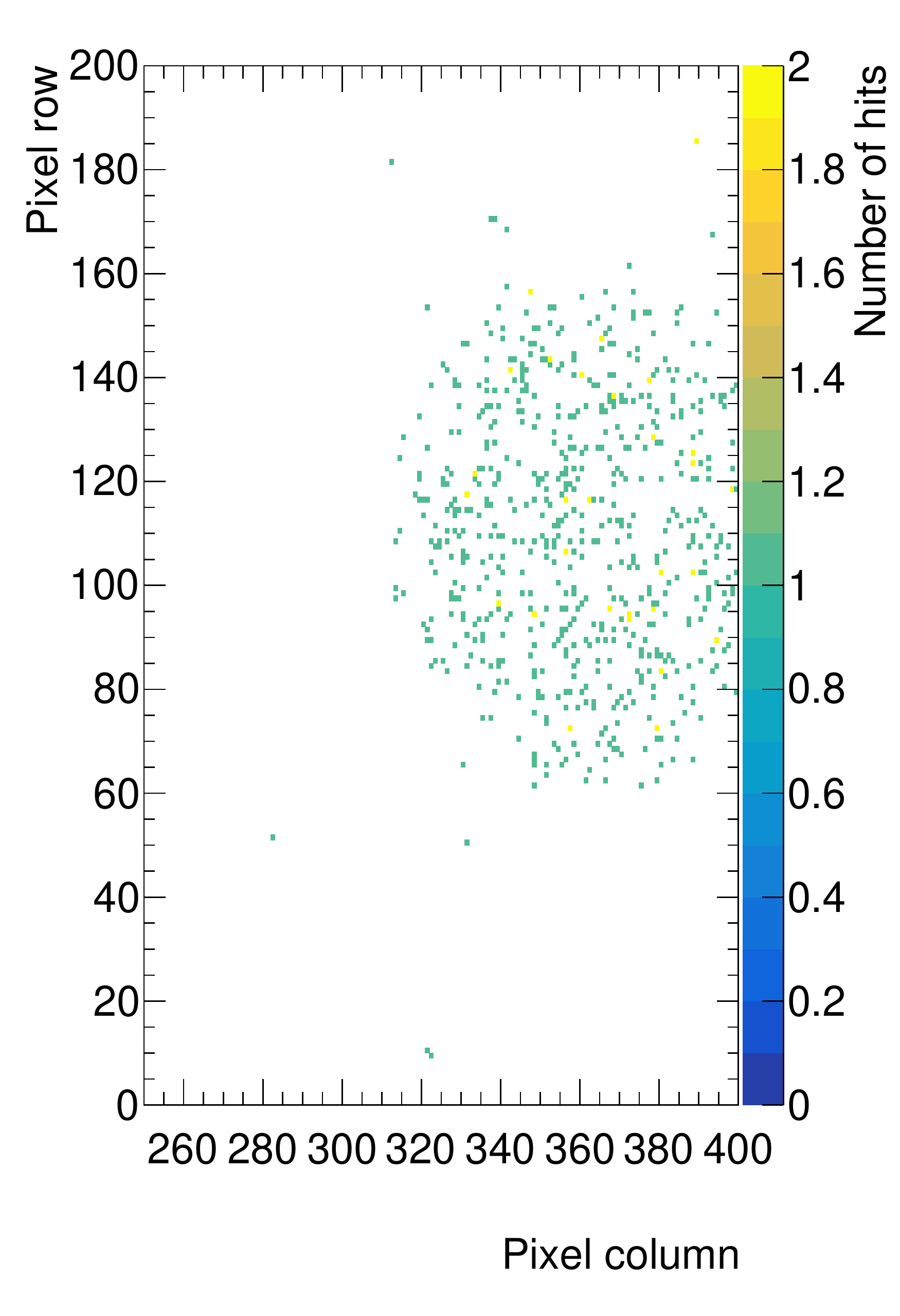}
  \caption{Spatial distribution of hits with ToT in the range [3,14] in the readout chip after a 100 hour run at 56$^\circ$.  The hits were associated with a Compton scattered photon detected by the spectrometer.  The majority of pixels had only 1 hit in the run, if they had any hit at all.}
  \label{hitmap}
\end{figure}

The spatial distribution (in readout chip row and column coordinates) of recorded hits after 100 hours of running at a 56$^\circ$ scattering angle is shown in~\ref{hitmap}.  These hits were read out based on a trigger caused by a photon being detected by the spectrometer, as illustrated in Figure~\ref{dataflow}.  The hits shown have ToT in the range [3,14] in order to show fewer noise and background hits.  The roughly circular distribution of hits is caused by the collimator, which was offset from the center of the module.  The majority of pixels received only 1 hit during the run, if they saw a hit at all.  4\% of hit pixels were hit twice throughout the run.  Similarly, 95\% of readout events had only one pixel activated.  In 0.4\% of events, two adjacent pixels were activated, creating a ``cluster''.  The remaining events had multiple non-adjacent pixel activated.  Such events are caused when a noise or background event is coincident with a Compton scatter.  The distribution of noise and background would mimic that discussed in Section~\ref{noise}.

\section{Example run and proof of principle}
\label{example}

\subsection{Test data}
\begin{figure}
  \centering
  \subfloat[]{\includegraphics[width=0.45\linewidth]{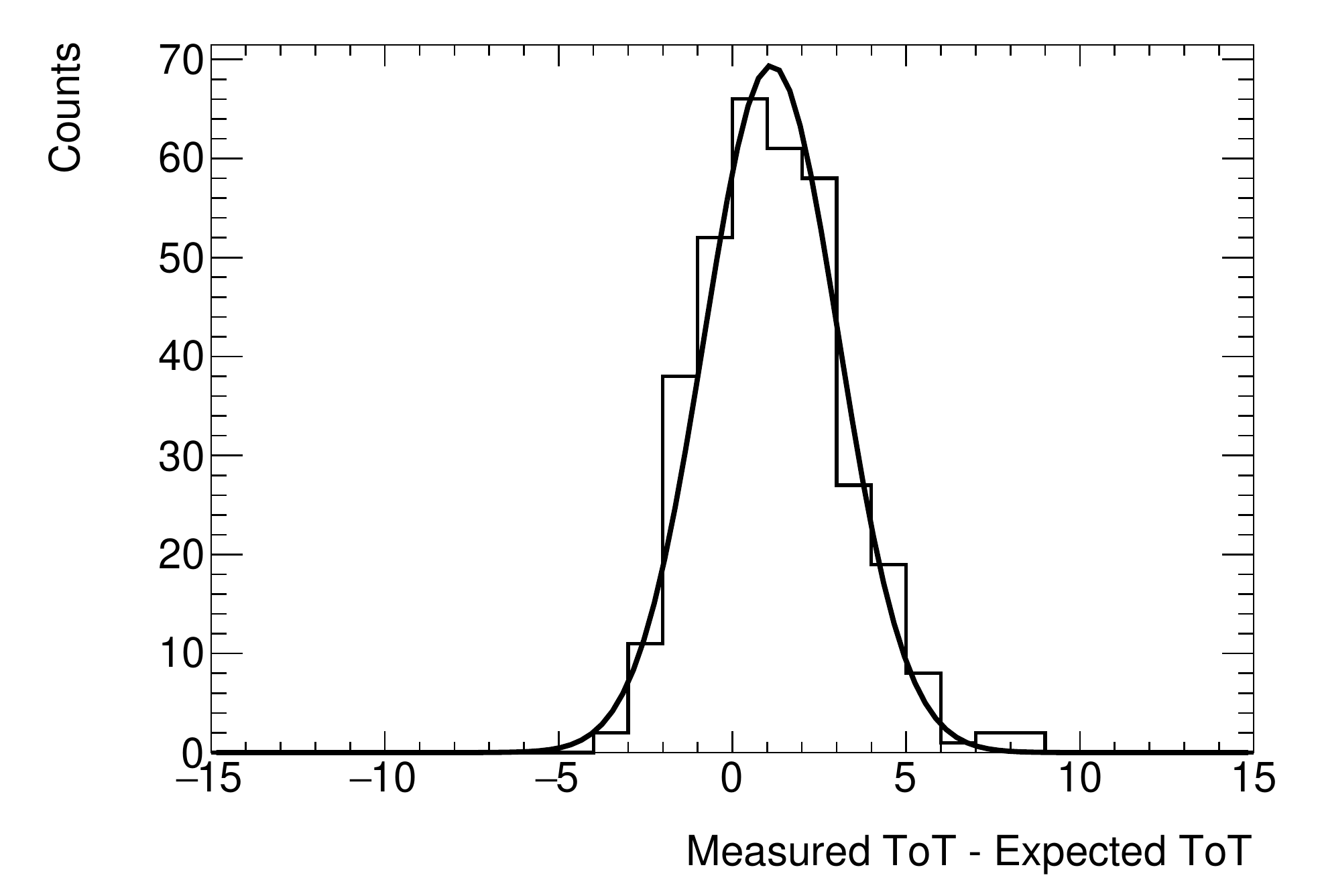}}
  \qquad
    \subfloat[]{\includegraphics[width=0.45\linewidth]{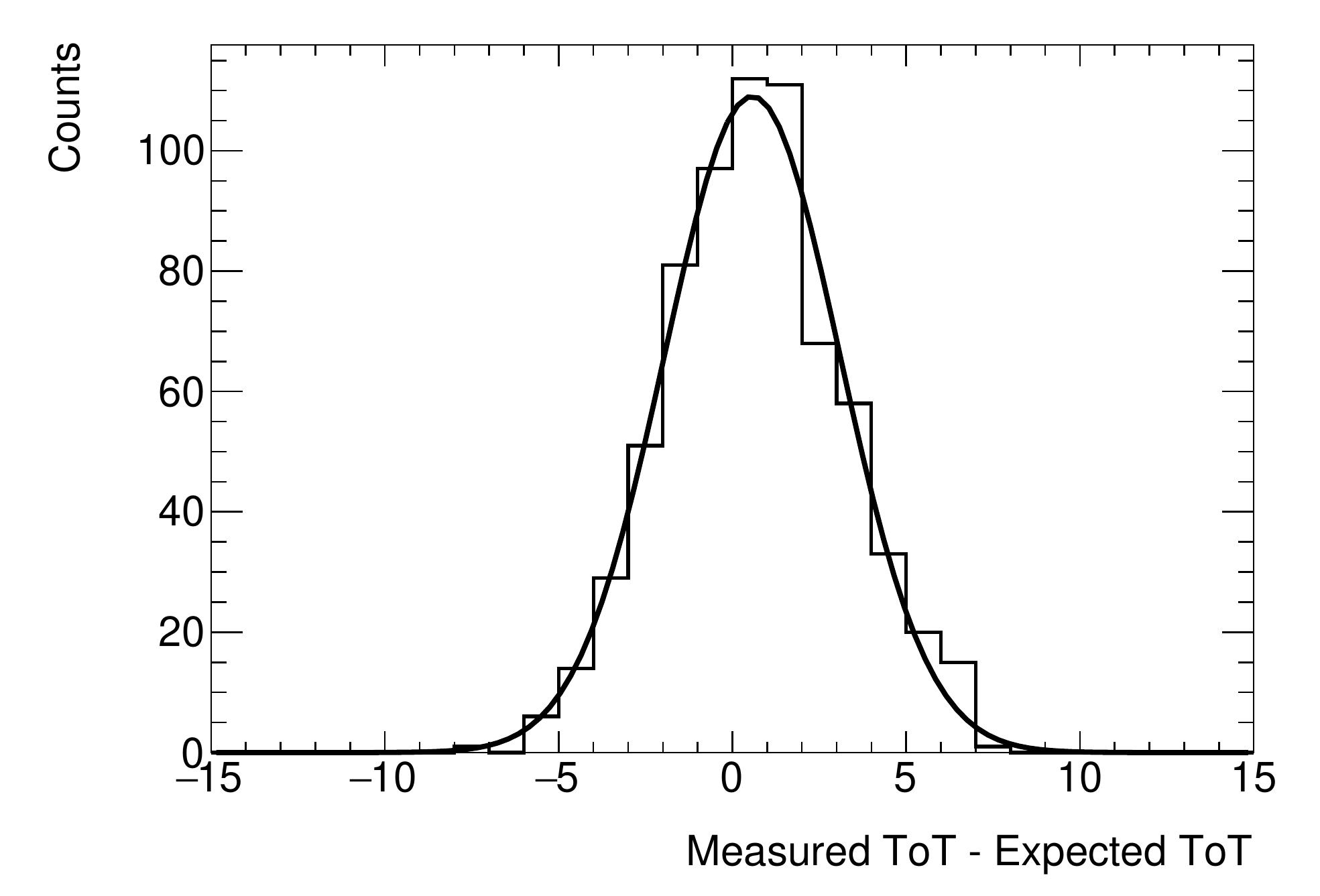}}
  \qquad
  \subfloat[]{\includegraphics[width=0.45\linewidth]{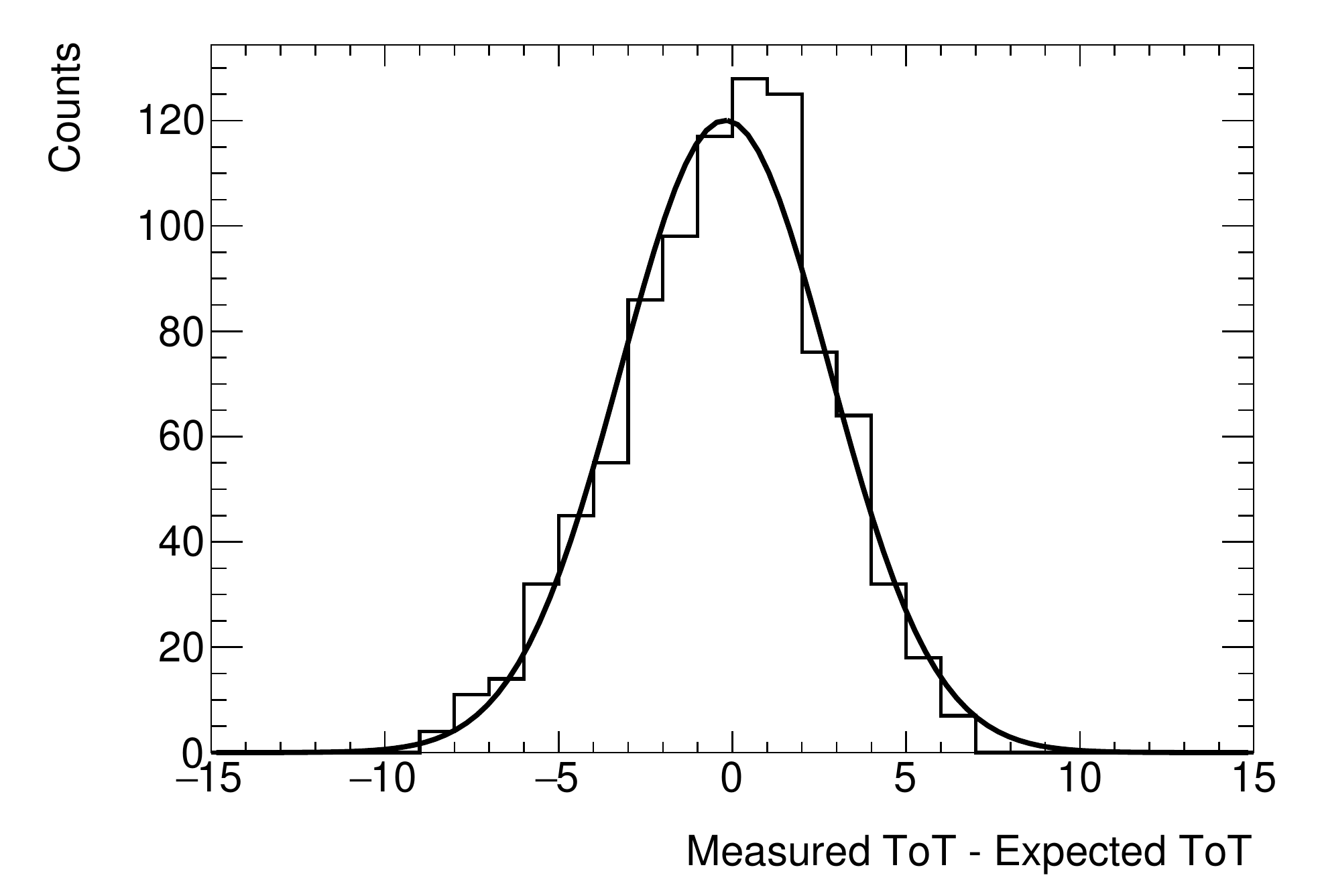}}
  \qquad
  \subfloat[]{\includegraphics[width=0.45\linewidth]{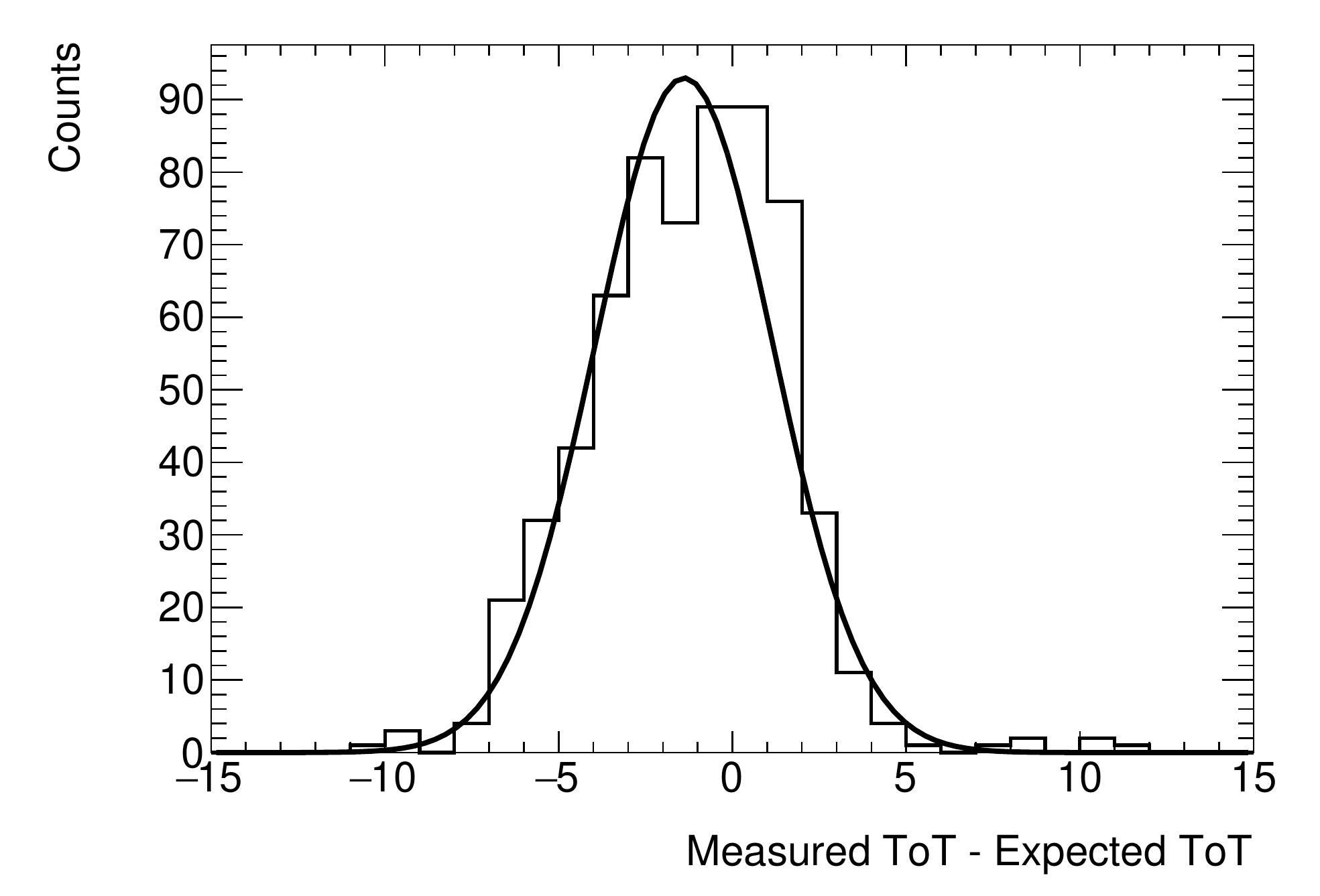}}
  \caption{Distributions of per-pixel differences between the ToT of hits from Compton scatters and the expected ToT.  The expected ToT is based on an internal injection of the charge deposit expected from from the Compton formula.  (a), (b), (c), and (d) show the data for the runs at 51$^\circ$, 56$^\circ$, 63$^\circ$, and 67$^\circ$, respectively.  Fits to a Gaussian function are also shown for each distribution.}
  \label{comparisons}
\end{figure}

To test the method, data-taking runs were performed at four scattering angles: 51$^\circ$, 56$^\circ$, 63$^\circ$, and 67$^\circ$.  Table~\ref{data_table} shows the expected energy of charge deposits and the numbers of hits with ToT in the range [4,13] for each of the runs.  Figure~\ref{comparisons} shows the distribution of per-pixel differences between the ToT of hits associated with Compton scattered photons and the average ToT expected from the calculated charge deposit (expected ToT).  Only hits with ToT in the range [4,13] are considered in order to reduce noise and backgrounds.  The expected ToT of each pixel is derived from fits such as those shown in Figure~\ref{totfunc}.  The results of fits to a Gaussian function are also plotted in Figure~\ref{comparisons}, and the mean and standard deviation results from the fits are presented in Table~\ref{data_table}.

\begin{table}[h!]
\caption{Summary and results of the four runs used to test the Compton-scattering calibration method.}
\centering
 \begin{tabular}{|c | c | c | c | c | c | c|} 
 \hline
 Angle [$^\circ$] & Deposited Energy [keV] & Charge [e-h pairs] & Number of hits & $\mu_{\mathrm{fit}}$ [clock cycles] & $\sigma_{\mathrm{fit}}$ [clock cycles] \\ [0.5ex] 
 \hline\hline
 51 & 2.5 & 690 & 347 & $1.1 \pm 0.11$ & $1.9 \pm 0.08$ \\ 
 \hline
 56 & 3.0 & 830 & 697 & $0.57 \pm 0.10$ & $2.5 \pm 0.07$ \\
 \hline
 63 & 3.7 & 1020 & 912 & $-0.21 \pm 0.10$ & $3.0 \pm 0.09$ \\
 \hline
 67 & 4.1 & 1140 & 630 & $-1.4 \pm 0.12$ & $2.6 \pm 0.07$ \\
 \hline
\end{tabular}
\label{data_table}
\end{table}


\section{Precision and results}
The Compton scattering calibration method is performed by comparing an observed ToT to an expected ToT.  The expected ToT is derived by using the expected mean external charge deposit as the argument in the charge to ToT response function.  The charge to ToT response was derived using the uncalibrated internal injection circuit.  Uncertainties on the function will result in uncertainty on the expected ToT.  A precise calibration using this method requires accurate knowledge of both the external charge deposit and the expected charge to ToT conversion functions.

The external charge deposit is dependent on the scattering angle, and can be derived using Equation~\ref{comp}.  Every deposit caused by scattering at the same angle is not expected to be the same, as the scattered photon energy distribution has a non-zero width.  However, the error on the centroid of this distribution comes from uncertainty on the angle.  Here, an uncertainty of $\pm 1^\circ$ is taken on the mean scattering angle.  The mean scattering angle is the angle of deflection of a photon that is emitted along the photon center-line and scatters into the center of the spectrometer in Figure~\ref{angles}.  The uncertainty comes from the measurement of the angle and translates to an uncertainty of 0.1 keV in deposited energy or 28 e-h pairs in deposited charge.

The finite width of the Compton scattering peak, as observed in Figure~\ref{spectrom_spectrum}, is unavoidable and means that the exact charge deposit in each pixel hit cannot be known precisely.  The width of the Compton scattering peak leads to the width of the distributions in Figure~\ref{comparisons}.  As noted in Section~\ref{trigger}, the FWHM of the Compton peak is 1.4 keV, which translates to a FWHM of 390 e-h pairs.  Based on knowledge of the charge to ToT conversions functions, such a FWHM should translate into a widths of about 3 ToT units in the the distributions in Figure~\ref{comparisons}.  Further, the ToT measurement itself has imperfect resolution.  As seen in Figure~\ref{single_inject}, the injection of a single charge value can result in a ToT distribution with FWHM of up to 3 ToT units, which is primarily caused by noise due to the small charge scale.  A combination of the energy width and ToT resolution lead to an expected standard deviation in the (observed - expected) distribution of about 2 ToT units, depending on the pixels involved.  The standard deviations seen in table~\ref{data_table} are slightly larger, but this is likely due to background hits, which will have an overall broadening effect.  An uncertainty of 0.5 clock cycles will be taken on the centroids in order to account for potential biases the background introduces.

The fits used as the charge to ToT response functions have errors of 5\%.

Combining the uncertainty on the Gaussian centroid, the uncertainty due to backgrounds, and the uncertainty from the charge to ToT response functions yields a conservative uncertainty of $\pm$~0.6 ToT unit on the expected ToT.

\begin{figure}[!htb]
  \centering
  \includegraphics[width=0.45\linewidth]{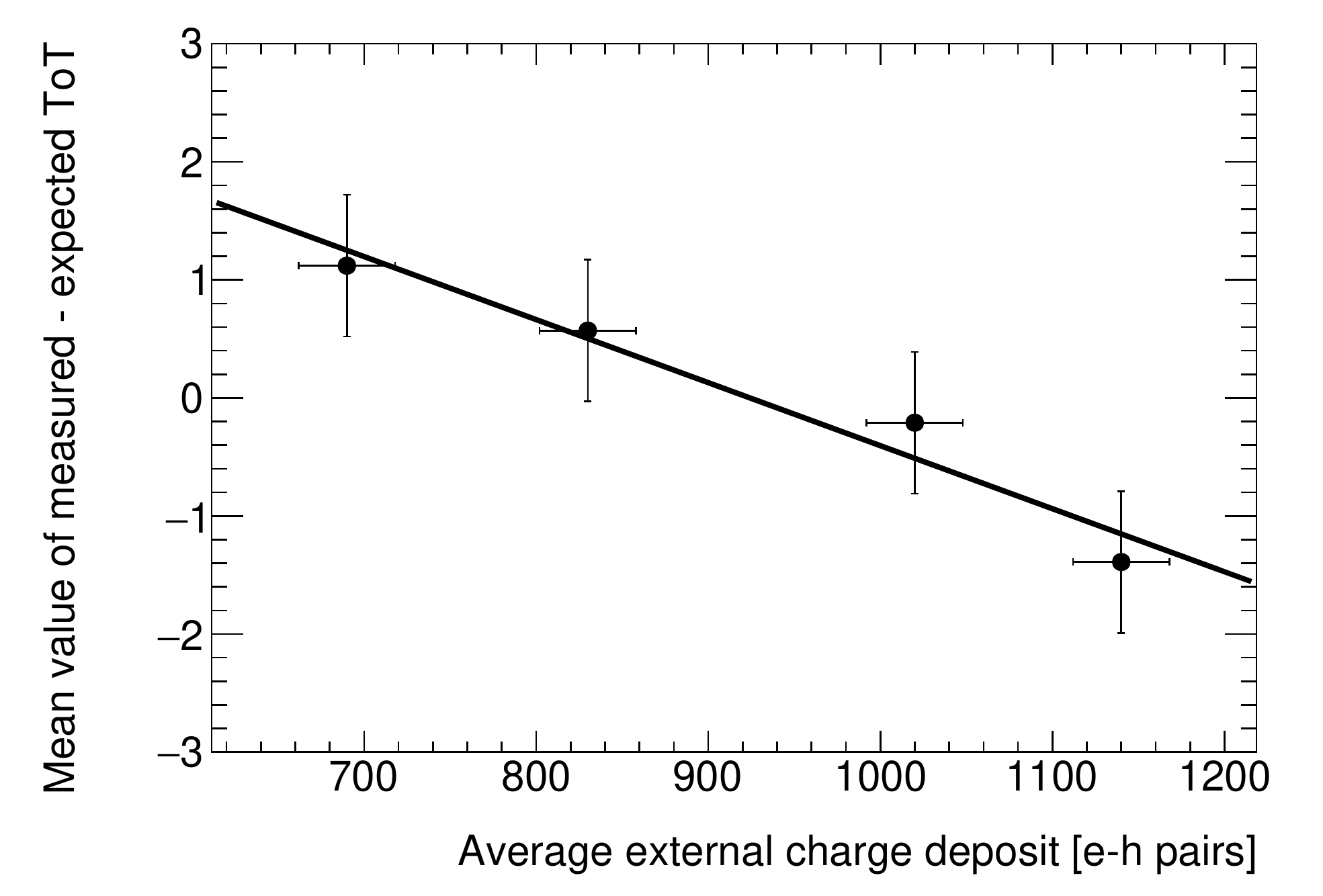}
  \caption{Plot of the mean values of the measured - expected ToT distributions found in Figure~\ref{comparisons} as a function of the expected charge deposit.  A linear trend line has been included with the four data points.}
  \label{diffplot}
\end{figure}

The mean values of the distributions in Figure~\ref{comparisons} are plotted against the expected charge deposit in Figure~\ref{diffplot}, along with the uncertainties on both the expected ToT and the average external charge deposit.   A fit to line was performed, finding a slope of $-0.53 \pm 0.18$ units of ToT per 100 e-h pairs difference in charge deposit for this particular tuning.  An intercept of $4.9 \pm 1.7$ ToT units was found, which translates to a difference of $2.5 \pm 0.2$ units of ToT at the threshold injection of 440 e-h pairs.

\section{Discussion and conclusions}
We have presented here a new method for the calibration of thin silicon sensors with energies in the range of 1 - 6 keV using the Compton scattering of photons.  The method was developed and tested using a 150 $\mu$m sensor bump-bonded onto an RD53A and an $^{241}$Am source.  The physical setup involved a spectrometer that can be easily repositioned to enable calibrations at any desired energy within the achievable range.  An example calibration was demonstrated, finding a trend in the difference between the ToT response to the external charge deposit and corresponding internal charge injection.

In the four data taking runs used for the example calibration, about 5 Compton-scattering hits were observed per hour.  That means that about 7\% of the pixels exposed by the collimator were hit during a 100 hour run.  Because of this, the four runs presented above were actually comprised of hits in largely non-overlapping sets of pixels.  Based on the current setup, about 700 hours of running would be needed to expect 50\% of pixels to be hit.  However, this required time could be reduced by adjusting the setup.  The hit rate will increase linearly with source activity.  The trigger rate will increase quadratically with distance from module to spectrometer (linearly with solid angle subtended), but this would lead to further broadening of the Compton scattering peak or create multiple peaks.  Instead, using multiple spectrometers (or a segmented spectrometer) will multiply the rate accordingly without loss of resolution, and will allow taking data for multiple energies at the same time.  In any case, data taking periods of at least a week long are likely needed in order to achieve adequate statistics.  This method is thus not suitable as a quality control tool to be applied to a large number of modules, but rather as a means to characterize the response and absolute charge scale on a single representative device.

Compared to the Compton scattering method, traditional calibration methods provide superior speed.  They also typically involved narrower energy peaks and smaller backgrounds.  However, these methods are best applied above 6 keV.  The Compton scattering calibration method provides access to a continuous spectrum of energies between 1 and 6 keV.  The setup and materials are readily available in most labs and cost less than an X-ray fluorescence setup.  An extension of this method could be to trigger on the Thompson peak instead of the Compton peak of Figure~\ref{spectrom_spectrum}, in order to reach very low energy deposits (of order eV and below) for experiments measuring low energy phonon signals instead of charge.

\section{Acknowledgements}
This work was supported by the U.S. Department of Energy, Office of Science under contract DE-AC02-05CH11231.  Patrick McCormack was supported by the National Science Foundation Graduate Research Fellowship under Grant No. DGE 1752814. Any opinion, findings, and conclusions or recommendations expressed in this material are those of the author and do not necessarily reflect the views of the National Science Foundation.

\bibliographystyle{unsrt}
\bibliography{references}  

\end{document}